\documentclass[fleqn,usenatbib]{mnras}

\usepackage{newtxtext,newtxmath}

\usepackage[T1]{fontenc}

\DeclareRobustCommand{\VAN}[3]{#2}
\let\VANthebibliography\thebibliography
\def\thebibliography{\DeclareRobustCommand{\VAN}[3]{##3}\VANthebibliography}

\usepackage{graphicx}
\usepackage{amsmath}
\usepackage{lineno}
\usepackage{subfigure}
\usepackage{ragged2e}

\graphicspath{{./}{figures/}}
\newcommand{\GP}{{\scshape GALPROP}}
\newcommand{\hess}{{H.E.S.S.}}
\newcommand{\tibet}{Tibet~AS${\gamma}$}
\newcommand{\fermi}{\textit{Fermi}--LAT}
\newcommand{\gray}{$\gamma$-ray}

\newcommand{\hi}{H\,{\scshape i}}
\newcommand{\htwo}{H$_2$}
\newcommand{\hii}{H\,{\scshape ii}}

\title[Diffuse TeV \gray{} Predictions with \GP]{The Steady-State Multi-TeV Diffuse Gamma-Ray Emission Predicted with \GP{} and Prospects for the Cherenkov Telescope Array}

\author[P. D. Marinos et al.]{
P. D. Marinos,$^{1}$\thanks{E-mail: peter.marinos@adelaide.edu.au}
G. P. Rowell,$^{1}$
T. A. Porter,$^{2}$
and G. Jóhannesson$^{3}$
\\
$^{1}$School of Physical Sciences, University of Adelaide, Adelaide, South Australia 5000, Australia\\
$^{2}$W. W. Hansen Experimental Physics Laboratory and Kavli Institute for Particle Astrophysics and Cosmology, Stanford University, Stanford, CA 94305, USA\\
$^{3}$Science Institute, University of Iceland, IS-107 Reykjavik, Iceland
}

\date{Accepted XXX. Received YYY; in original form ZZZ}

\pubyear{2022}

\begin{document}
\label{firstpage}
\pagerange{\pageref{firstpage}--\pageref{lastpage}}
\maketitle

\begin{abstract}
    Cosmic Rays~(CRs) interact with the diffuse gas, radiation, and magnetic fields in the interstellar medium~(ISM) to produce electromagnetic emissions that are a significant component of the all-sky flux across a broad wavelength range.
    The \textit{Fermi} Large Area Telescope~(LAT) has measured these emissions at GeV \gray{} energies with high statistics.
    Meanwhile, the High-Energy Stereoscopic System~(\hess) telescope array has observed large-scale Galactic diffuse emission in the TeV \gray{} energy range.
    The emissions observed at GeV and TeV energies are connected by the common origin of the CR particles injected by the sources, but the energy dependence of the mixture from the general ISM~(true `diffuse'), those emanating from the relatively nearby interstellar space about the sources, and the sources themselves, is not well understood.
    In this paper, we investigate predictions of the broadband emissions using the \GP{} code over a grid of steady-state 3D~models that include variations over CR sources, and other ISM target distributions.
    We compare, in particular, the model predictions in the VHE~($\gtrsim$100\,GeV) \gray{} range with the \hess{} Galactic plane survey~(HGPS) after carefully subtracting emission from catalogued \gray{} sources.
    Accounting for the unresolved source contribution, and the systematic uncertainty of the HGPS, we find that the \GP{} model predictions agree with lower estimates for the HGPS source-subtracted diffuse flux.
    We discuss the implications of the modelling results for interpretation of data from the next generation Cherenkov Telescope Array~(CTA).
\end{abstract}




\section{Introduction}

Cosmic ray particles are injected by sources, propagating over millions of years and eventually pervading the Galaxy.
This process results in a `sea' of CRs that produce broadband all-sky emissions due to energy losses with the other components of the diffuse ISM: the interstellar gas, the interstellar radiation field~(ISRF), and the Galactic magnetic field~(GMF).
The leakage into the ISM from the individual CR sources also produces localised enhancements of the particle intensities on top of the background `sea', which also contribute to the broadband non-thermal sky brightness.
The diffuse emissions by the CRs therefore encode critical information on their sources, how they are injected into the ISM and their propagation history, as well as the spatial distributions of the other ISM components.
Observations of their non-thermal emissions can be used to provide essential insights for understanding how the CRs are accelerated up to the highest energies within the Galaxy.

At GeV \gray{} energies the sky is dominated by the emissions produced by the CR sea throughout the Milky Way~(MW), which have been measured and studied extensively with the \fermi~\citep{AckermannM.2012}.
Recovering individual CR source characteristics at these energies relies on extracting them from the much brighter diffuse fore-/background emissions that they are embedded within.
Determination of these fore-/background emissions using a physically motivated methodology is desirable as it enables a more reliable estimate of the true source characteristics.
Furthermore, residuals resulting from the subtraction of the fore-/background emissions can be interpreted as true features of physical interest.
Generally, analyses of the \fermi{} data employ physically motivated interstellar emissions models~(IEMs) based on gas tracers and other templates modelled using the \GP{} code~\citep[e.g.][]{AceroF.2016,2016ApJ...819...44A,2016ApJS..224....8A}.
Systematic studies using such IEMs have been employed for extracting source properties, in particular, for various supernova remnants~(SNRs) that are the putative major CR source class~\citep{2016ApJS..224....8A}.

For the VHE range, the emissions about source regions are brighter than the diffuse emission.
The recent release of the \hess{} Galactic plane survey~\citep{HESS_GPS.2018} shows localised, extended regions embedded in lower-intensity, broadly distributed emissions -- similar to but not exactly the same as the \fermi{} data at lower energies.
This is due to combined effects.
The VHE CRs are more concentrated about the sources.
Also, \hess{} uses `off source' fields for background estimation, which likely removes some diffuse emissions and does not enable probing lower fluxes.
Recovering source properties for dimmer sources requires proper assessment and accounting for the diffuse emissions.

The VHE diffuse emission is currently observed with low significance by \hess{}~\citep{AbramowskiA.2014}, and the spectral extrapolation of a \fermi{} IEM indicates a general compatibility with the TeV flux observed by ARGO-YBJ~\citep{BartoliB.2015}.
There is a connection between the GeV and TeV energy ranges, but the energy dependence of the mixture of emissions from the general ISM~(true `diffuse'), those emanating from the relatively nearby interstellar space about the sources, and the sources themselves, is not well understood.
Looking forward, accurately determining their relative contributions will be essential for the next generation of facilities that will have a significantly enhanced sensitivity.
An example is the Cherenkov Telescope Array~\citep[CTA;][]{TheCTAOConsortium.2018}, which is currently under construction.

CTA will have a much larger field of view~($\sim$10$^\circ$) than \hess, an improved angular resolution~(by a factor $\gtrsim$~5), and will be at least an order of magnitude more sensitive~\citep{2011ExA....32..193A}.
The lower flux levels that CTA will reach, and its much larger field of view, will make it extremely sensitive to the details of the diffuse emissions.
CTA will also detect hundreds of new \gray{} sources.
Consequently, source confusion, together with the other emissions, will likely be a significant issue that will need to be addressed.
Distinguishing between the diffuse and general ISM related emissions, and extended emission coming from the $\gtrsim$10s--100s~pcs region surrounding the sources, will be critical for accurate recovery of the embedded source properties~\citep{DubusG.2013, ACERO2013276, AmbrogiL.2016}.

In this paper, we use the \GP{} CR propagation package to model the diffuse emission into the VHE range for a grid of models that are categorised by CR source density and ISM target density distributions.
The models are all spatially 3D and the CR intensity distributions through the ISM are solved for the steady-state case with the propagation parameters optimised to reproduce local CR data.
We compare the model predictions with the HGPS observations, carefully accounting for the survey characteristics.
We investigate the variation of the predictions for the diffuse emissions across this model grid and use it to estimate the modelling uncertainty.
The \GP{} predictions are comparable to lower limits on the diffuse emission at TeV energies inferred from the HGPS observations, when accounting for current estimates of the unresolved source fraction in the data.
We discuss also how the accurate modelling of the diffuse emissions and creation of TeV IEMs will be a critical element for maximising the scientific return from the more sensitive observations by facilities such as CTA.%
\section{\GP{} Modelling} \label{sect:GALPROP}

The \GP{} framework~\citep{1998ApJ...509..212S,1998ApJ...493..694M} is a widely employed CR propagation package that now has over 25~years of development behind it.
For this paper, we use the latest release~(v57), where an extensive description of the current features is given by~\citet{2022ApJS..262...30P}.

\subsection{Model Setup}

For an assumed propagation phenomenology, the critical inputs for a \GP{} run are the CR source density distribution, together with the distributions of the other ISM components that result in the energy losses and corresponding secondary particle and emissions production.
The CR source distribution is subject to considerable uncertainty, and we construct a representative set of 3D models varying the respective content of sources distributed in the Galactic disc and spiral arms.
Meanwhile, \GP{} has well developed 3D models for the interstellar gas and radiation fields, and a collection of GMF models taken from the literature.
We fix the interstellar gas model and make predictions for combinations of the interstellar radiation and magnetic field distributions over the CR source density distributions.
We describe in detail below these elements for our calculations.

\subsubsection{The Interstellar Gas}

The ISM gas consists mostly of H and He with a ratio of 10:1 by number~\citep{2001RvMP...73.1031F}.
It can be found in the different states, atomic~(\hi), molecular~(\htwo), or ionised~(\hii), while He is mostly neutral.
\hi{} is $\sim$60\% of the mass, while \htwo{} and \hii{} contain 25\% and 15\%, respectively \citep{2001RvMP...73.1031F}.
The \hii{} gas has a low number density and scale height $\sim$few~100\,pc.
The \htwo{} gas is clumpy and forms high density molecular clouds.

For this paper, we use the 3D gas density distributions for the neutral ISM developed by~\citet{JohannessonG.2018}.
They were obtained using a maximum-likelihood forward model folding method to the LAB \hi-survey~\citep{KalberlaP.2005} and the CfA composite CO survey~\citep{DameT.2001}.
These neutral gas models include a warped and flaring disc, four spiral arms, and a central bulge.
For the \hii{} we employ the NE2001 model~\citep{2002astro.ph..7156C, 2003astro.ph..1598C, 2004ASPC..317..211C} with the updates by~\citet{2008PASA...25..184G}.
The CR energy losses and fragmentation~(for nuclei) depend on the spatial density distribution for the interstellar gas.

For calculating gas-related \gray{} intensities we use the gas column-density maps based on the HI4PI survey~\citep{HI4PI.2016} and the composite CO survey, with a correction for the contribution of missing so-called `dark gas'~\citep{2005Sci...307.1292G} using a map of optical depth at 353\,GHz based on {\it Planck} data~\citep{Planck.2016}.
The gas column density maps for atomic and molecular hydrogen are split into Galactocentric radial bins using a rotation curve~\citep{AckermannM.2012} due to the limited kinematic resolution of the data.
The line-of-sight integration uses the gas density model to accurately weight the CR flux over the extent of the individual rings.
This enables the use of the full resolution of the CR intensity solutions that determine the secondary emission emissivities.
Helium and heavier elements are assumed to be identically distributed and are accounted for by correction factors.

\subsubsection{The CR Source Distributions} \label{ssect:GALPROP Sources}

While the SNRs are believed to be the principal CR source class, pulsar wind nebulae~(PWNe) and other types likely contribute at some level, but the relative amount is poorly known.
For the purpose of this paper, we do not specify the individual source classes that are typically given according to 2D Galactocentric averaged distributions~\citep[e.g.][]{AckermannM.2012}.
Instead we construct a set of synthetic distributions with systematic gradation of the proportion between disc and spiral arm components to investigate the effect of the 3D source distribution on the VHE \gray{} emissions.

We follow~\citet{PorterT.2017} and~\citet{JohannessonG.2018} using the disc-like distribution from~\citet{2004A&A...422..545Y} and four spiral arms.
The spiral arms have the geometry of those for the R12 ISRF model~(see below), but with equal weighting for their normalisations.
Our CR source distributions start with a purely disc-like distribution~(that we term SA0), and increase with relative contribution by the spiral arms until the distribution is purely due to the spiral arms~(termed SA100).
Consequently, we have the SA0, SA25, SA50, SA75, and SA100 distributions corresponding to 0, 25, 50, 75, and 100\%, respectively, for the source luminosity contained in the spiral arms, with the remaining source luminosity in the disc-like component.
The primary CR source spectra and other parameters are determined for each model by the optimisation procedure described below.

\subsubsection{The Interstellar Radiation and Magnetic Fields}

The CR electrons and positrons lose energy via Compton and synchrotron interactions with the interstellar radiation and magnetic fields.
Into the VHE range, these processes strongly influence the CR spectral intensities and correspondingly affect the intensity distribution for the $\gamma$~rays.
Because there is still uncertainty for the ISRF and GMF distributions, we employ representative models that are available within the \GP{} framework.

The ISRF encompasses the electromagnetic radiation within the Galaxy, including emission from stars, infrared light from interstellar dust radiating heat, and the cosmic microwave background~(CMB).
The state-of-the-art 3D ISRF models for the MW were developed by~\citet{PorterT.2017} based on spatially smooth stellar and dust models.
They have designations R12 and F98 that correspond to the respective references supplying the stellar/dust distributions~\citep{RobitailleT.2012,FreudenreichH.1998}.
They similarly reproduce the data, but neither is an overall best match.
The R12 model provides better correspondence toward the spiral arm tangents, but does not display the asymmetry associated with the bar~(R12 has an axisymmetric stellar bulge), and the stellar disc scale length is incompatible with the near-IR profiles.
Meanwhile, the F98 model has the disc scale-length in better agreement with the near-IR data, incorporates the bulge/bar asymmetry, but has none of the structure associated with the spiral arms.

Both R12 and F98 models are providing equivalent solutions for the ISRF distribution.
At least toward the inner Galaxy, where the ISRF intensity is most uncertain, they are providing lower and upper bounds as determined by pair-absorption effects on sources toward the Galactic centre~(GC)~\citep[][]{2018PhRvD..98d1302P}.

The GMF consists of the large-scale regular~\citep{1985A&A...153...17B} and small-scale random~\citep[e.g.][]{2008A&A...477..573S} components that are about equal in intensity.
The random fields are mostly produced by the SNe and other outflows, which result in randomly oriented fields with a typical spatial scale of $\lesssim$100\,pc~\citep{1995MNRAS.277.1243G, 2008ApJ...680..362H}.
Also, there may be the anisotropic random~(`striated') fields that are a large-scale ordering originating from stretching or compression of the random field~\citep{2001SSRv...99..243B}.
This component is expected to be aligned to the large-scale regular field, with frequent reversal of its direction on small scales.
\GP{} includes multiple large-scale MW GMF models~\citep{2008A&A...477..573S, 2010RAA....10.1287S, PshirkovM.2011, 2012ApJ...757...14J, 2010MNRAS.401.1013J}.

In this paper we use the bisymmetric spiral GMF model from~\citet{PshirkovM.2011}, that we refer to as PBSS, as a representative model with a spiral structure. We also employ the \GP{} axisymmetric exponential distribution~\citep[GASE;][]{StrongA.2000}.
It has the functional form for Galactocentric coordinates $R,Z$:

\begin{eqnarray}
    B(R, Z) = B_{0} \exp \left( -\frac{R-R_{0}}{H_{R}} \right) \exp \left( -\frac{|Z|}{H_{Z}} \right) \label{eq:GASE}
\end{eqnarray}

\noindent
where $B_{0}=5\,\mu \mathrm{G}$ is the magnetic field strength, $R_{0}=8.5$\,kpc is the IAU recommended distance from the GC to Earth~\citep{1986MNRAS.221.1023K}, and $H_{R}=10$\,kpc and $H_{Z}=2$\,kpc are the scale heights for the radial and height components, respectively.
This distribution is a simple, exponentially decreasing field in both Galactocentric radius and height above the plane, and does not include the spiral arms.
Although there are more modern GMF distributions that have better agreement with radio data, including those mentioned above, the GASE GMF is still commonly used throughout the literature~\citep[e.g.][]{2022PhRvD.105j3033K, 2022FrPhy..1744501Q}. The use of the GASE GMF in this work is therefore to allow us to better characterise the variation between commonly used GMF distributions.

\subsubsection{Parameter Optimisation}

For this paper we assume a diffusive-reacceleration propagation model with isotropic and homogeneous spatial diffusion coefficient\footnote{Note that the spatial diffusion coefficient for the CR propagation can be linked to the GMF strength distribution in \GP~\citep{2015ApJ...799...86A}, but we do not use this functionality for the present paper.}.
The consistency condition is that each combination of the inputs described above reproduces the local CR spectra.
To ensure this, we optimise the propagation parameters and source spectra as described below.

The CR propagation calculations use a 3D right-handed spatial grid with the solar system on the positive $X$-axis and $Z=0$\,kpc defining the Galactic plane with $R_S = 8.5$\,kpc for the distance from the Sun to the GC.
For runtime efficiency we use the non-uniform grid functionality included with the v57 \GP{} release~\citep{2022ApJS..262...30P} for solving the propagation equations.
We employ the tangent grid function where the parameters for the transformation function are chosen so that the $X/Y$ resolution nearby the solar system is $\sim$30\,pc, increasing to $\sim$1.2\,kpc at the boundary of the Galactic disc, which is at 20\,kpc from the GC.
In the $Z$-direction the resolution is 15\,pc in the plane, increasing to 0.6\,kpc at the boundary of the grid at $|Z_{\rm halo}| = 6$\,kpc~\citep{2016ApJ...824...16J}.
The kinetic energy grid is logarithmic from 1\,GeV to 1\,PeV with ten planes per decade.

We follow the procedure in \citet{PorterT.2017} and~\citet{JohannessonG.2018}, with the CR data from~\citet{JohannessonG.2019}~(see their table~1).
For each of the source distributions, an initial optimisation of the propagation models is made by fitting to the observed CR spectra from AMS-02 and \textit{Voyager~1} in the GeV energy range, where the CR sea is the dominant source of CRs. This procedure is performed for the CR species: Be, B, C, O, Mg, Ne, and Si. These are kept fixed, and the injection spectra for electrons, protons, and He are tuned together. As the proton spectrum impacts the normalisation of the heavier species, and therefore also impacts their propagation parameters, this process is performed iteratively until convergence. As we are extrapolating outside the energy range of the data, the best-fit model for the source distributions is determined via a $\chi^{2}$ test, and then its local CR spectra are used to re-optimise the CR data for the other source distributions. This is to ensure that all five source distributions give the same local CR spectra, reducing inconsistences between the distributions due to limited data statistics and coverage over the modelled energy range.
Examples between the model and data agreement are provided in~\citet{PorterT.2017} up to 1\,TeV/nucleon. For higher energies, the spread of the observations is larger, and a range of spectral indices can be consistent with the measurements. For example, for protons with energies around $10^{6}$\,GeV the difference between experiments in the measured fluxes with uncertainties is approximately an order of magnitude~\citep{2022icrc.confE..94C}.
The parameters that vary with the source distributions are shown in Table~\ref{tab:SA pars}.

The CR injection spectra are given by the power-law $\mathrm{d}n/\mathrm{d}p \propto R^{-\gamma}$ with two spectral breaks, and is defined separately for electrons, protons, Helium, and a single spectral form for all CR species heavier than Helium. The proton and electron spectra are normalised at the Solar position to $J_{p}$ and $J_{e^{-}}$ at the energies $E_{\mathrm{kin}, p}=100$\,GeV and $E_{\mathrm{kin}, e^{-}}=34.5$\,GeV respectively, with heavier elements normalised relative to the proton spectrum at 100\,GeV/nuc. The CR spectral breaks are given by the rigidities $R_{1, \mathrm{CR}}$ and $R_{2, \mathrm{CR}}$, and the spectral indices are given by $\gamma=\gamma_{0, \mathrm{CR}}$ for rigidities $R<R_{1, \mathrm{CR}}$, $\gamma=\gamma_{1, \mathrm{CR}}$ for rigidities $R_{1, \mathrm{CR}}<R<R_{2, \mathrm{CR}}$, and $\gamma=\gamma_{2, \mathrm{CR}}$ for rigidities $R>R_{2, \mathrm{CR}}$. The diffusion coefficient is given by the power-law $D(R) \propto \beta R^{\delta}$, and is normalised to $D_{0, xx}$ at the rigidity $R_{0, D}=4$\,GV with the spectral index given by $\delta=\delta_{0}$. The normalisation rigidity, spectral breaks, and spectral index after the break are held constant across all models. Finally, as the \gray{} spectrum at a given energy always depends on more energetic CRs, we set the nuclei cut-off energy~($E_{\mathrm{max}}$) to 1\,PeV/nuclei across all models to ensure the correct treatment of the \gray{} spectrum up to 100\,TeV.

Although there are hints for a cut off in the local~(i.e.~post propagation) electron spectrum above $\sim$10\,TeV~\citep[e.g.][]{2009A&A...508..561A,2017Natur.552...63D}, local measurements are unable to constrain the CR electron spectrum across the entire MW due to the short~($<$1\,kpc) travel distances of multi-TeV electrons. As electrons are known to be accelerated in the MW up to PeV energies by PWNe, no artificial cut off is applied to the CR electron injection spectrum for our models.

\begin{table}
    \centering
    \caption{The optimised \GP{} propagation parameters for each of the five different source distributions.}
    \label{tab:SA pars}
    \begin{tabular}{lccccc}
        \hline
        Parameter & SA0 & SA25 & SA50 & SA75 & SA100 \\
        \hline
        $D_{0,xx}$~[$10^{28}$]        & 4.36  & 4.39  & 4.55  & 4.67  & 4.66  \\ 
        $\delta_{0}$                  & 0.354 & 0.349 & 0.344 & 0.340 & 0.339 \\ 
        $v_{\mathrm{Alfven}}$         & 17.8  & 18.2  & 18.1  & 19.8  & 19.1  \\ 
        $J_{p}$~[$10^{-9}$]           & 4.096 & 4.404 & 4.113 & 4.329 & 4.394 \\ 
        $J_{e^{-}}$~[$10^{-10}$]      & 3.925 & 4.444 & 3.994 & 4.428 & 4.502 \\ 
        $\gamma_{0,e^{-}}$            & 1.616 & 1.390 & 1.488 & 1.455 & 1.521 \\ 
        $\gamma_{1,e^{-}}$            & 2.843 & 2.756 & 2.766 & 2.763 & 2.753 \\ 
        $\gamma_{2,e^{-}}$            & 2.493 & 2.460 & 2.470 & 2.447 & 2.422 \\ 
        $R_{1,e^{-}}$                 & 6.72  & 5.27  & 5.14  & 5.54  & 5.29  \\ 
        $R_{2,e^{-}}$                 & 52.4  & 81.6  & 67.7  & 70.7  & 79.7  \\ 
        $\gamma_{0,\mathrm{p}}$       & 1.958 & 1.928 & 1.990 & 1.965 & 2.009 \\ 
        $\gamma_{1,\mathrm{p}}$       & 2.450 & 2.464 & 2.466 & 2.494 & 2.481 \\ 
        $\gamma_{2,\mathrm{p}}$       & 2.391 & 2.411 & 2.355 & 2.374 & 2.414 \\ 
        $R_{1,\mathrm{p}}$            & 12.0  & 12.3  & 12.2  & 14.5  & 13.5  \\ 
        $R_{2,\mathrm{p}}$            & 202   & 157   & 266   & 108   & 125   \\ 
        $\gamma_{0,\mathrm{He}}$      & 1.925 & 1.886 & 1.956 & 1.937 & 1.971 \\ 
        $\gamma_{1,\mathrm{He}}$      & 2.417 & 2.421 & 2.432 & 2.467 & 2.443 \\ 
        $\gamma_{2,\mathrm{He}}$      & 2.358 & 2.369 & 2.320 & 2.347 & 2.376 \\ 
        $R_{1,\mathrm{He}}$           & 12.0  & 12.3  & 12.2  & 14.5  & 13.5  \\ 
        $R_{2,\mathrm{He}}$           & 202   & 157   & 266   & 108   & 125   \\ 
        $\gamma_{0,\mathrm{Z}}$       & 1.328 & 1.519 & 1.426 & 1.630 & 1.624 \\ 
        $\gamma_{1,\mathrm{Z}}$       & 2.377 & 2.390 & 2.399 & 2.399 & 2.418 \\ 
        $\gamma_{2,\mathrm{Z}}$       & 2.377 & 2.390 & 2.399 & 2.399 & 2.418 \\ 
        $R_{1,\mathrm{Z}}$            & 3.16  & 4.21  & 3.44  & 4.61  & 4.50  \\ 
        $R_{2,\mathrm{Z}}$~[$10^{3}$] & 5.00  & 5.00  & 5.00  & 5.00  & 5.00  \\ 
        \hline
    \end{tabular}
    \\
    \justifying
    \noindent
    \footnotesize{$\gamma_{0, \mathrm{CR}}$ is the power-law index before the first break, $\gamma_{1, \mathrm{CR}}$ between the first and second, and $\gamma_{2, \mathrm{CR}}$ after the second break.} \\
    \footnotesize{$R_{1, \mathrm{CR}}$~[GV] is the rigidity of the first break, and $R_{2, \mathrm{CR}}$~[GV] is the rigidity of the second break.} \\
    \footnotesize{The diffusion coefficient~($D_{0,xx}$) is measured in cm$^{2}$\,s$^{-1}$, the Alfven velocity~($v_{\mathrm{Alfven}}$) in km\,s$^{-1}$, and the CR normalisation constants~($J_{p}$ and $J_{e^{-}}$) are measured in MeV$^{-1}$\,cm$^{-2}$\,s$^{-1}$\,sr$^{-1}$.}
\end{table}

\subsection{Interstellar Emissions Modelling Predictions} \label{ssect:GALPROP setup and results}

The propagation model parameters for each source distribution~(Table~\ref{tab:SA pars}) are used to calculate steady-state CR intensity solutions across the Galaxy using \GP.
The spatial grid used for the CR propagation calculations is also employed for determining \gray{} emissivities. 
The \gray{} intensity maps at the solar system location are obtained by line-of-sight~(LOS) integration of the \gray{} emissivities for the standard processes~($\pi^0$-decay, IC scattering), where the emissivities are determined for a logarithmic grid from 1\,GeV to 100\,TeV using five bins per decade spacing.
We use a HEALPix \citep{HEALPix} order 9 isopixelisation for the skymap generation\footnote{This gives a pixel size of $6.9^{\prime}$, which is similar to the point-spread function of \hess~($4.8^{\prime}$).}.
All calculations of the IC component use the anisotropic scattering cross section~\citep{2000ApJ...528..357M}, which accounts for the full directional intensity distribution for the R12/F98 ISRF models.
The $\gamma\gamma\rightarrow e^\pm$ attenuation that affects the \gray{} intensities mainly $\gtrsim$10\,TeV, is included in the LOS integration, where the optical depth calculation includes the directionality of the ISRF \citep{2006ApJ...640L.155M,2018PhRvD..98d1302P}.
For the R12/F98 ISRF models, the relevant precomputed optical depth maps included with the v57 \GP{} release are employed.

\begin{figure*}
    \centering
    \subfigure{
        \includegraphics[width=0.43\textwidth]{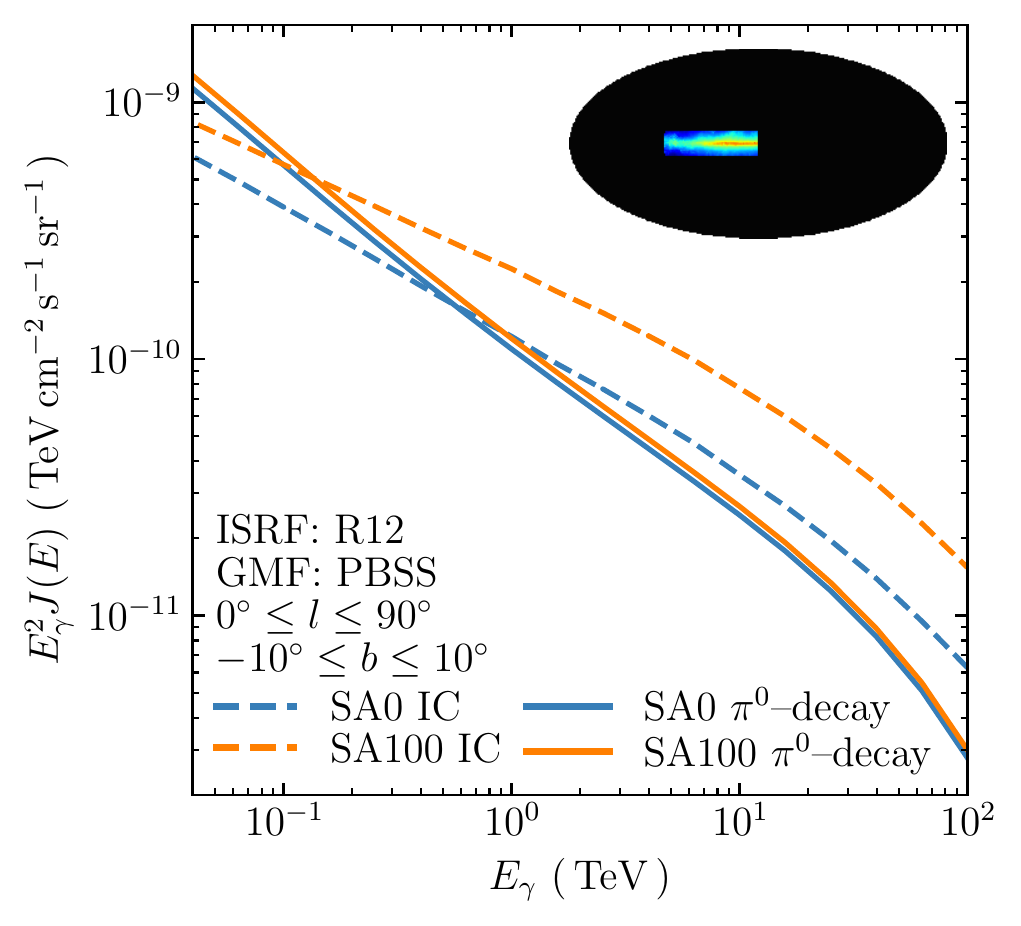}
        \includegraphics[width=0.43\textwidth]{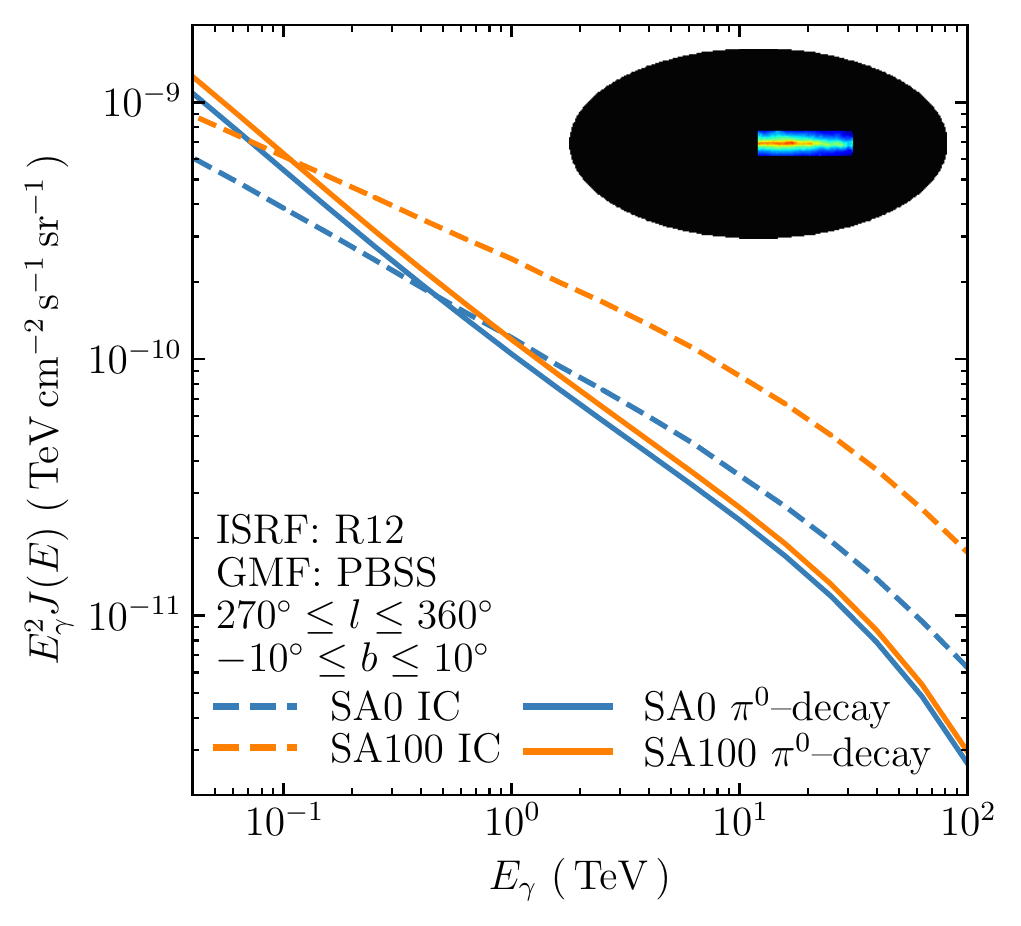}
    }
    \subfigure{
        \includegraphics[width=0.43\textwidth]{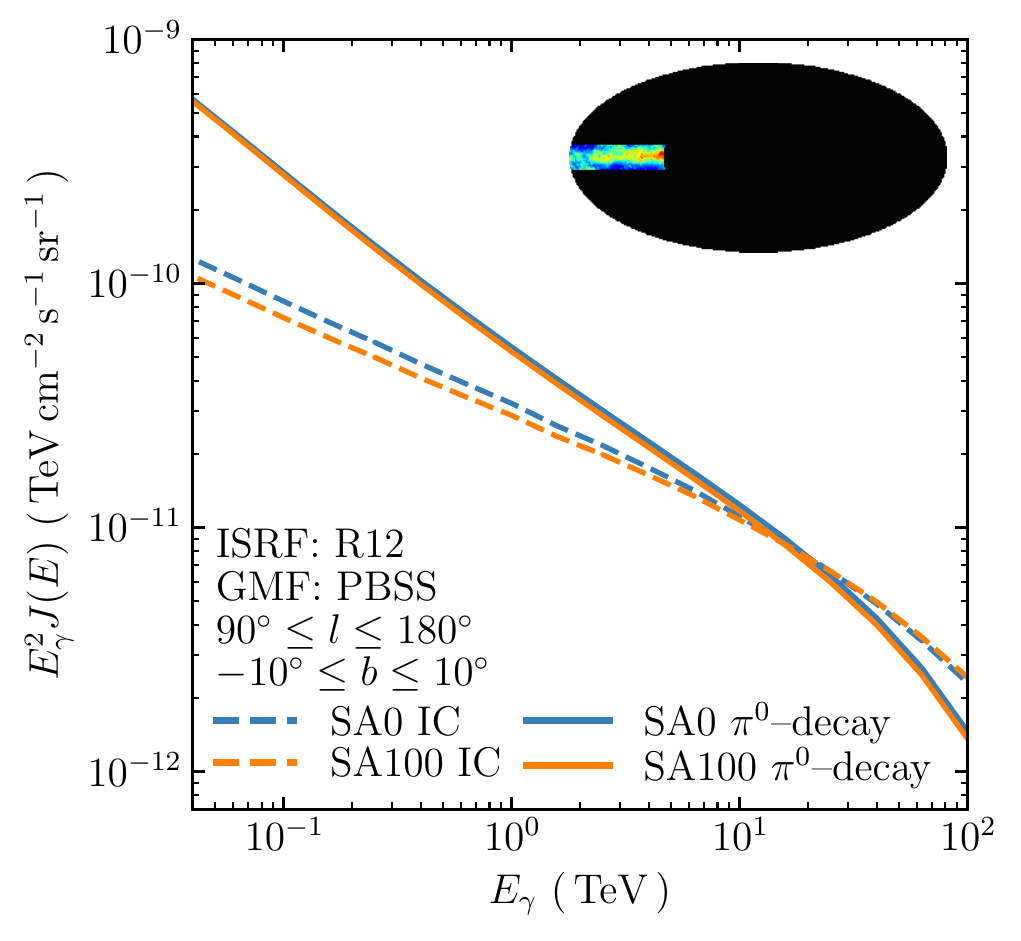}
        \includegraphics[width=0.43\textwidth]{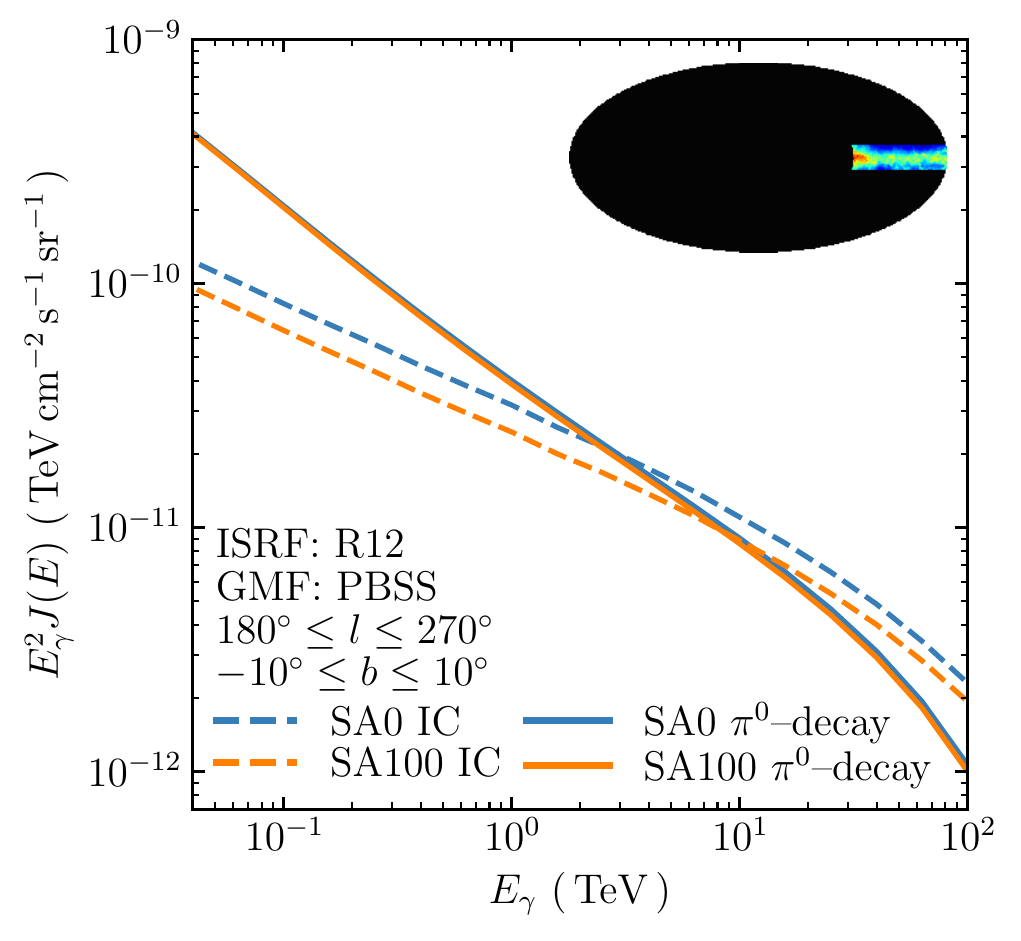}
    }
    \subfigure{
        \includegraphics[width=0.43\textwidth]{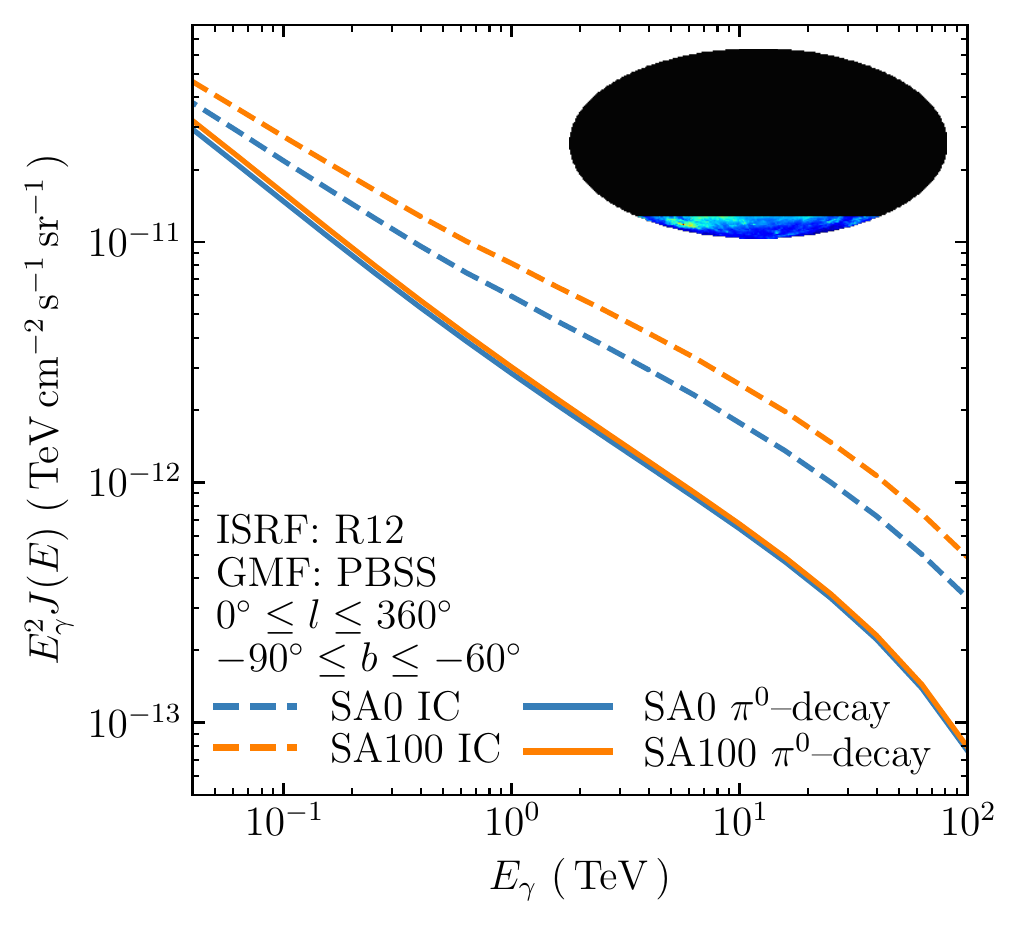}
        \includegraphics[width=0.43\textwidth]{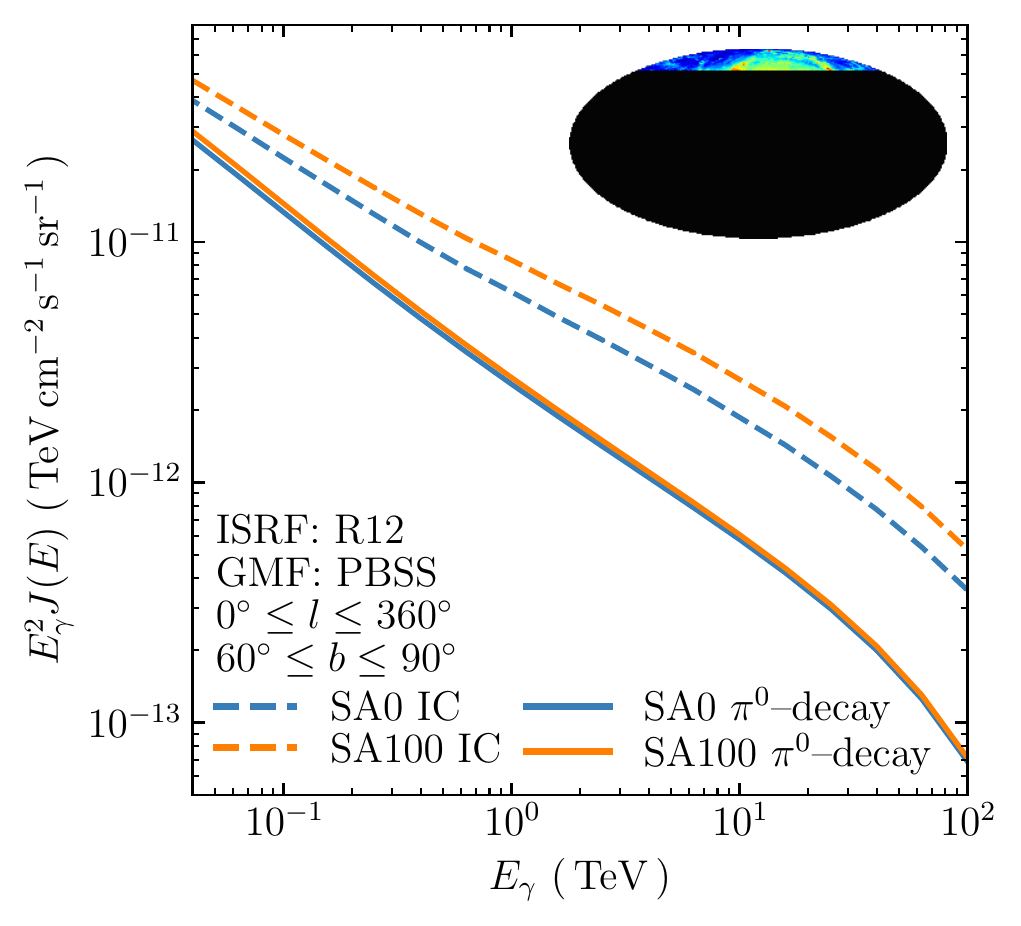}
    }
    \caption{Model spectra for the \gray{} emission for four quadrants along the Galactic plane as well as the poles, with the inset showing the sky regions from which the flux was taken. The SA0 and SA100 models are shown as upper and lower bounds, with the solid lines corresponding to $\pi^{0}$-decay and the dashed lines corresponding to IC emission. For all regions, the ISRF was chosen to be R12 and the GMF was chosen to be PBSS.}
    \label{fig:region fluxes}
\end{figure*}

Fig.~\ref{fig:region fluxes} shows the predicted fluxes from 1\,GeV to 100\,TeV averaged over different Galactic quadrants about the plane and within $30^\circ$ of the north/south polar regions.
For clarity, we show only the modelled fluxes for the SA0 and SA100 distributions for the R12/PBSS ISRF/GMF combination~(other combinations display qualitatively similar trends).
Along the plane, for the central longitudes~($l\leq|90^{\circ}|$) and central latitudes~($b\leq|10^{\circ}|$) the IC emission starts to account for a large fraction of the total \gray{} emission around 100\,GeV, and dominates the TeV Galactic emission after 10\,TeV, for all source distributions.
For the outer-Galactic region, where there are fewer CR sources~($l\geq|90^{\circ}|$), the IC provides a strong relative contribution until $\sim$10\,TeV energies, when there is less variation overall between the SA distributions.
Meanwhile, towards the polar regions~($b\geq|60^{\circ}|$) there is a much lower column density for the ISM gas, which results in a reduction of the $\pi^0$-decay emissions intensity.
Because the scale height of the ISRF is larger than that of the gas, the IC emission is reduced to a lesser degree over these parts of the sky.
This results in it being the strongest component of the \gray{} sky brightness for all SA distributions across the entire spectrum, from 1\,GeV to 100\,TeV.

\begin{figure}
    \centering
    \includegraphics[width=0.9\linewidth]{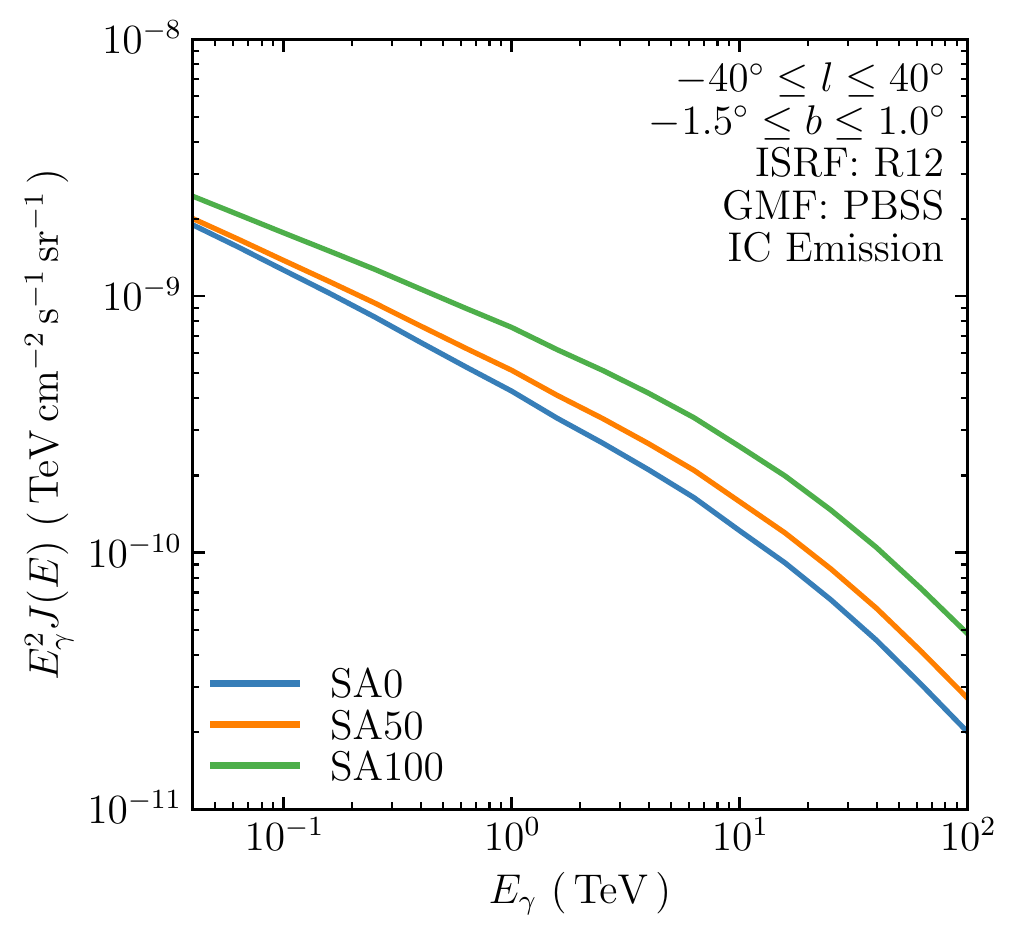}
    \includegraphics[width=0.9\linewidth]{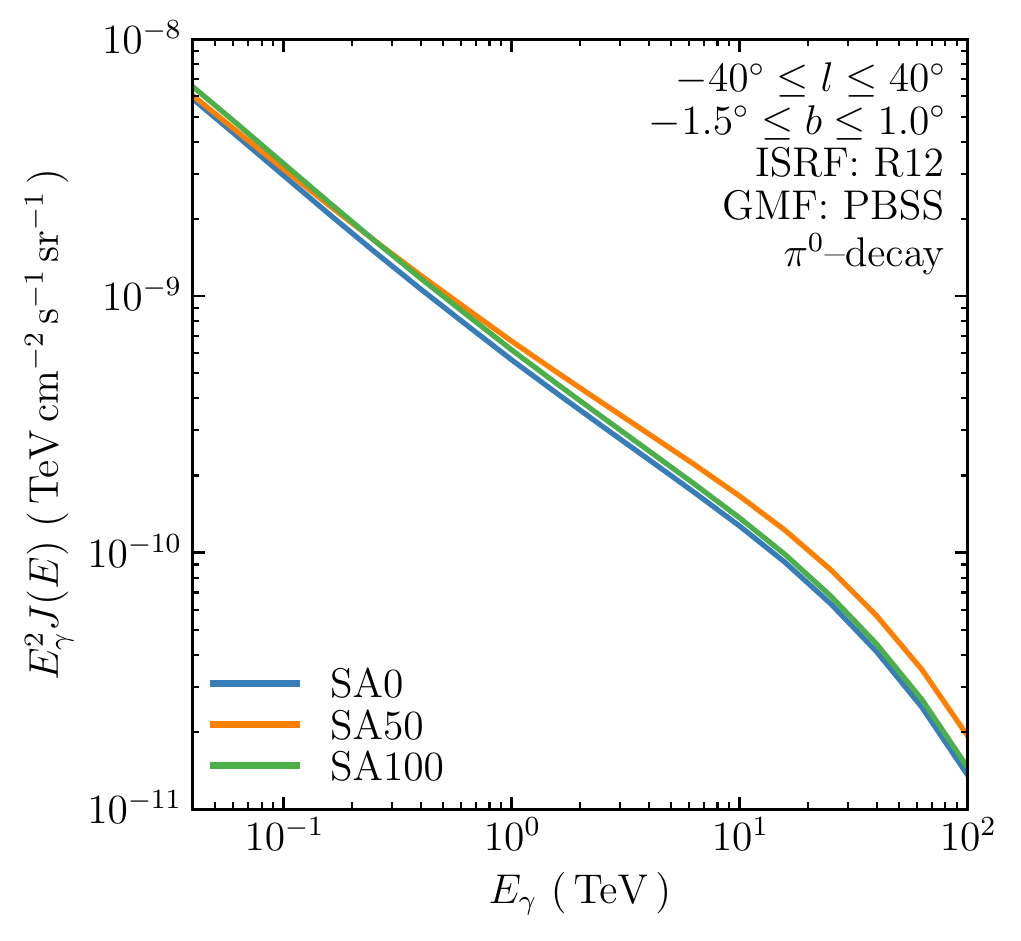}
    \caption{The model spectra averaged over the box~($l \leq |40^{\circ}|$) and~($-1.5^{\circ} < b < 1.0^{\circ}$) for the R12 ISRF and the PBSS GMF. SA0 is shown in blue, SA50 in orange, and SA100 in green. TOP: IC emission, BOTTOM: $\pi^{0}$-decay.}
    \label{fig:IC and pion flux}
\end{figure}

Fig.~\ref{fig:IC and pion flux} shows the total flux for the various SA distributions averaged over the HGPS analysis region~($l \leq |40^{\circ}|$ and $-1.5^{\circ} < b < 1.0^{\circ}$).
As the cooling time for VHE CR electrons is short, the impact of the source distribution is most easily seen for the IC emission contribution.
Source density distributions with a higher weighting for CR sources in the arms generally boost the IC signal.
This is due to the greater concentration of CR injection about the arms together with the higher ISRF/GMF densities that are also similar in distribution to the sources.
This `density squared' effect~\citep{PorterT.2017} enhances the emissions over the more smoothly distributed model combinations.
The impact is such that for the SA100 distribution the IC flux at 100\,TeV is a factor~$\sim$2 brighter than that from the SA0 distribution.

\begin{figure}
    \centering
    \includegraphics[width=0.9\linewidth]{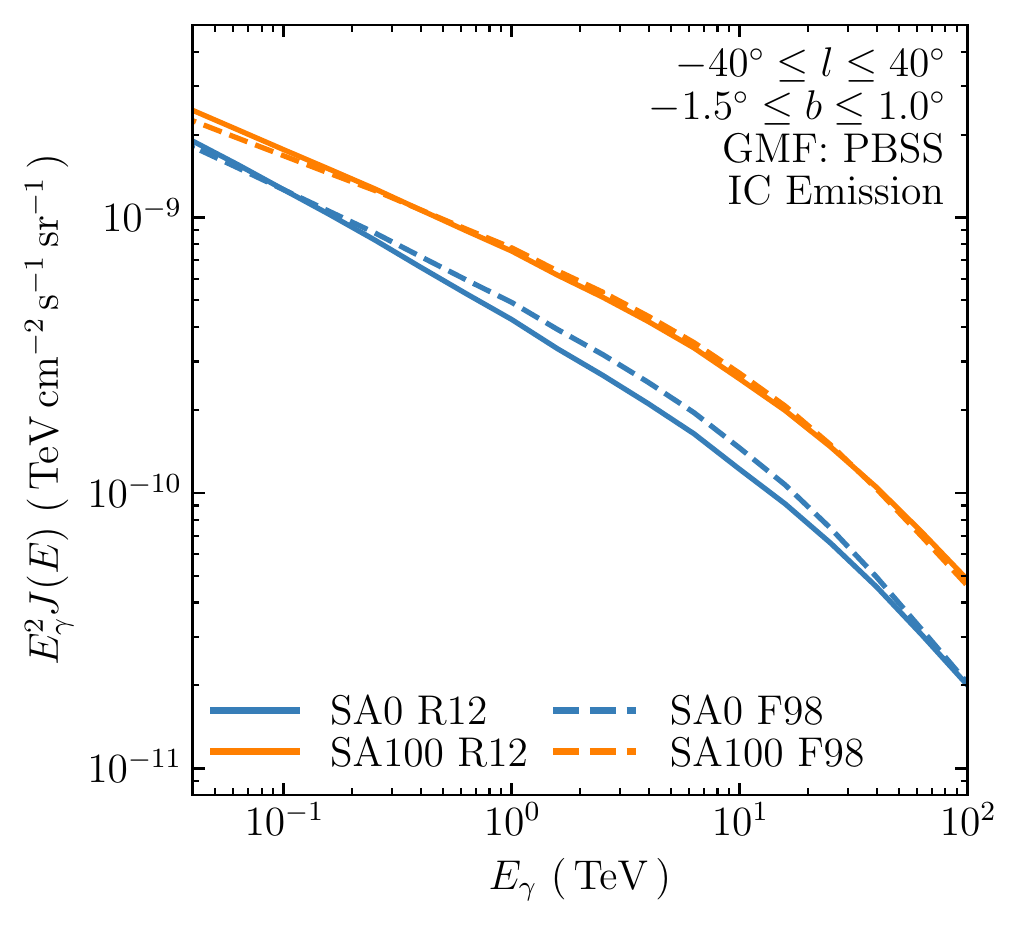}
    \caption{The model spectra of the IC emission averaged over the box~($l \leq |40^{\circ}|$) and~($-1.5^{\circ} \leq b \leq 1^{\circ}$) for the PBSS GMF. The source distributions shown are SA0~(blue) and SA100~(orange), and the ISRF distributions shown are R12~(solid line) and F98~(dashed line).}
    \label{fig:ISRF flux}
\end{figure}

Enhancements due to the ISRF distribution mainly affect the IC emissions for energies $\lesssim$10\,TeV.
In Fig.~\ref{fig:ISRF flux} we show predicted fluxes over the HGPS region with variation for the ISRF model.
Below $\sim$100\,GeV the stronger optical ISRF about the spiral arms in the R12 model enhances the IC emissions when compared to the F98 model.
At higher energies the Klein-Nishina~(KN) suppression reduces this effect, and the predictions for both become closer due to how the IR component of the respective ISRF models is distributed relative to the CR electrons.
For energies $\sim$1--2\,TeV the IC emission for the SA0 distribution varies $\sim$10\% between the R12 and F98 ISRFs, while there is little variation switching between the ISRF models for the SA100 distribution.
This variation for the SA0 distribution is due to the peaking of the source density about the GC.
There is a correspondingly higher electron intensity across the inner Galaxy that can produce IC emissions.
While for the SA100 distribution there is little variation as the inner cut-off for the spiral arms effectively produces a central `hole' in the electron intensity distribution.
Therefore, there is much less IC emission coming from the inner Galaxy for this source density distribution.
For energies $\gtrsim$10\,TeV even the IR component becomes KN supressed as well, and only scattering of the CMB is contributing to the emission.
The result is a negligible difference between the ISRF models at higher energies, where the different predictions is due solely to the varying source density distributions.

\begin{figure}
    \centering
    \includegraphics[width=0.9\linewidth]{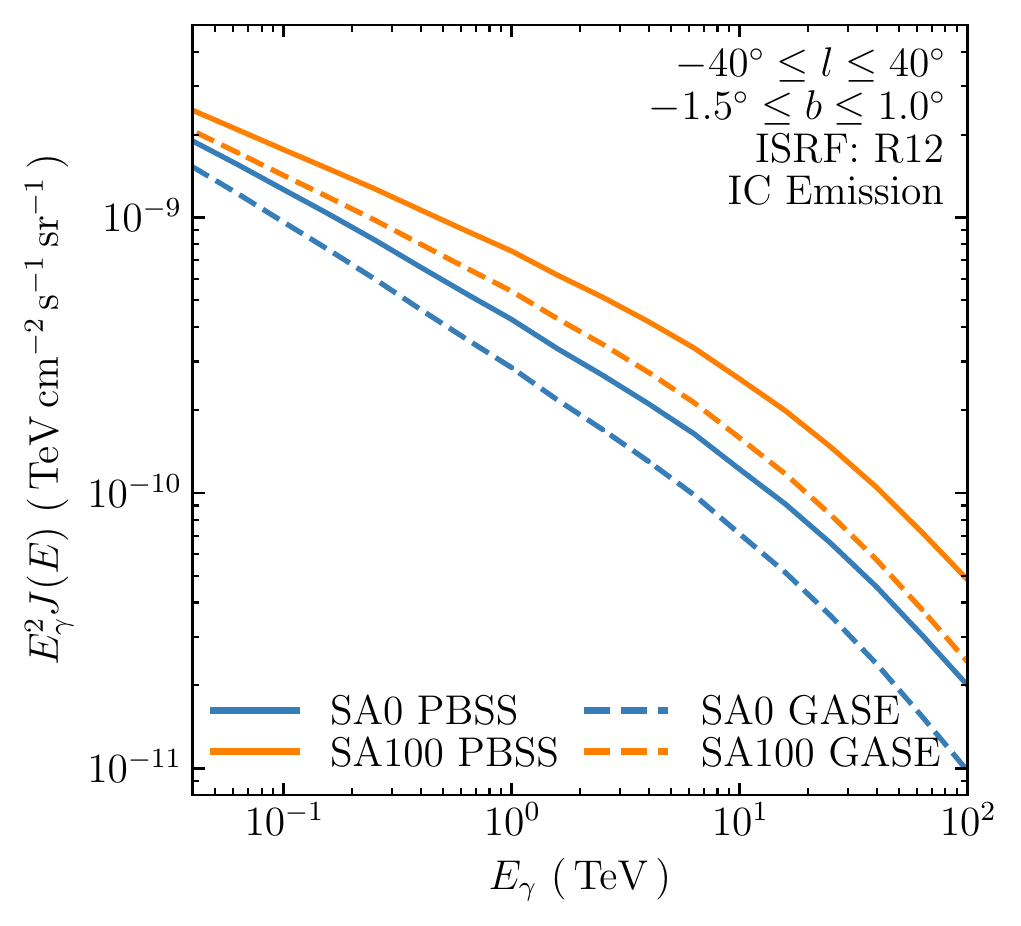}
    \caption{The model spectra of the IC emission averaged over the box~($l \leq |40^{\circ}|$) and~($-1.5^{\circ} \leq b \leq 1^{\circ}$) for the R12 ISRF. The source distributions shown are SA0~(blue) and SA100~(orange), and GMF distributions shown are PBSS~(solid line) and GASE~(dashed line).}
    \label{fig:GMF flux}
\end{figure}

Fig.~\ref{fig:GMF flux} shows the variation of the IC flux when the GMF model is varied for fixed ISRF distribution.
Varying the GMF distribution has a significant effect on the IC emission as energy increases into the VHE range due to the synchrotron losses dominating the cooling time scale for the electrons.
At 1\,TeV the difference in the IC emission between the PBSS and GASE distributions is approximately 30\%, increasing to 100\% by 100\,TeV. For synchrotron emission, the energy radiated by TeV electrons over time~($\dot{E} / E$) is proportional to the square of the magnetic field strength~($B$) of the medium that they diffuse through and the energy of the particle, i.e.~$\dot{E} / E \propto E^{2} B^{2}$. As the radiative losses of electrons depends strongly on the magnetic field strength, regions with higher GMF energy density will rapidly be depleted of high energy electrons. The GASE distribution has the most intense region of the field at the GC, leading to the IC emission being greatly reduced in the GC when compared to the PBSS distribution.

\begin{figure}
    \centering
    \includegraphics[width=\linewidth]{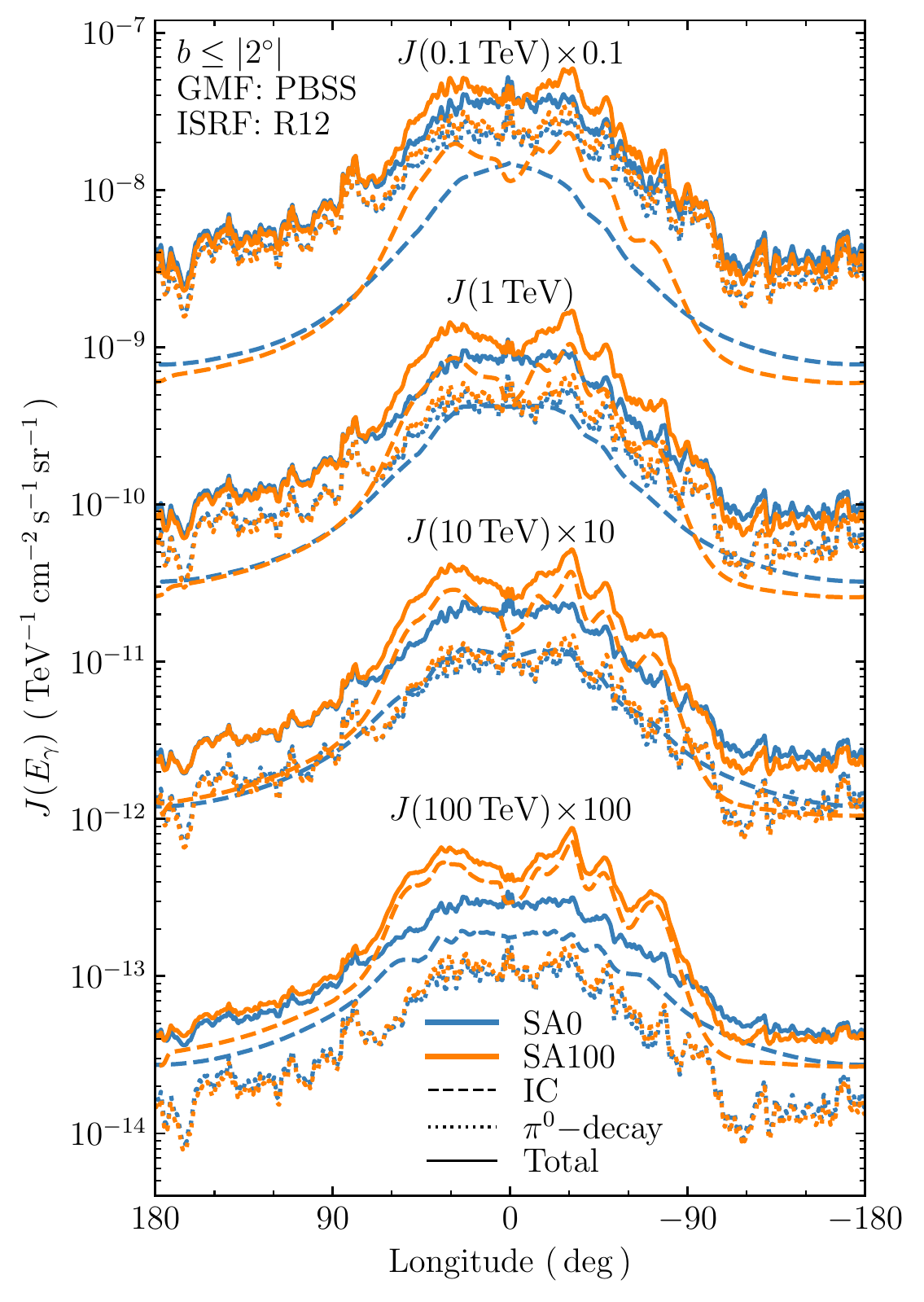}
    \caption{Longitude profile for $b \leq |2^{\circ}|$ along the Galactic plane for various energies using the R12 ISRF and the PBSS GMF distributions. The source distributions shown are SA0~(blue) and SA100~(orange), and the emission types shown are IC~(dashed), $\pi^{0}$-decay~(dotted), and the total flux~(solid). The profiles have been multiplied by the given factors to increase contrast.}
    \label{fig:diff. profile}
\end{figure}

Fig.~\ref{fig:diff. profile} shows the longitudinal profile of the IC and $\pi^{0}$-decay emission for the various source distributions for energy levels between 0.1\,TeV and 100\,TeV.
The $\pi^{0}$-decay dominates the \gray{} emission for all longitudes for the energies $E <$1\,TeV.
Because the energy losses for CR nuclei are slow and the diffusion is relatively fast, the propagated distributions for them are fairly smoothly distributed over all source distributions.
There is, correspondingly, minimal impact on the $\pi^0$-decay component in the longitudinal profiles.
Meanwhile, the IC emission becomes dominant for $\gtrsim$10\,TeV energies.
Because the energy losses are much faster for the electrons, its contribution to the profiles is much more sensitive to changes in the source distribution over most of the Galactic plane.
As the majority of the spiral arms are located between Earth and the GC, with one of the arms located in close proximity to the Earth, we see that for a larger fraction of CRs being injected into the spiral arms~(e.g.~SA100) the IC component becomes more intense for the central region~(i.e.~$l \leq |50^{\circ}|$).
The IC emission is also boosted along the spiral arm tangents located at $l\approx |20^{\circ}|$.
For the negative longitudes between the anti-centre and perpendicular to the GC, i.e.~for $-90^{\circ} < l < -180^{\circ}$, the IC emission decreases with an increasing fraction of CR injection in the spiral arms, and the difference between the source density distributions increases with increasing energy.

Across the entire plane for all source distributions the IC emission dominates the \gray{} emission at 100\,TeV.
It is not a surprise that the spiral structure in the source density distributions including arms is evident in the profiles.
However, at these energies there also appears to be structure for the SA0 density distribution IC profile similar to those models including arms that is not apparent at lower energies.
Because of the KN suppression for the optical and IR components of the ISRF, the only photon field is the~(uniform) CMB for IC processes.
The energy density of the GMF varies, but is much higher than that for the CMB about the spiral arms, and hence determines the energy loss time scale affecting the electron distribution.
The observed structure in the 100\,TeV IC profile for the SA0 distribution is therefore entirely due to that encoded in the PBSS GMF distribution, with its influence imprinted on the electron energy density.

For the spiral arms in the source distributions, the closest arm in the sky region $-90^{\circ} < l < -180^{\circ}$ is the Perseus arm located $\sim$3\,kpc away.
For the source distributions with a high fraction of CR injection in the spiral arms~(e.g.~SA75 and SA100), there is a large void of electron sources between the Solar System and the Perseus arm, resulting in reduced TeV IC emission in that direction.
However, as the Galactic disc component of the source distributions is smoothly varying, the source distributions with a low fraction of CR injection in the spiral arms~(e.g.~SA0 and SA25) have boosted TeV IC emission in these regions, as CRs are injected in the voids between the Solar System and the Perseus arm.
This is not seen in the positive longitudes far from the GC~(~$90^{\circ} < l < 180^{\circ}$) as the Perseus arm is close enough to the Solar System such that there is no void in the electron densities in any of the source distributions.
The proximity of the Perseus arm leads to a lower degree of variation in the IC emission in this region of the Galactic plane across the SA distributions.
Figs.~\ref{fig:GMF flux} and~\ref{fig:diff. profile} therefore show that the source distributions with a larger fraction of sources in the spiral arms show increased IC emission along the arm tangents, with the IC emission dominating over the pion-decay at 1\,TeV along the arms.

\begin{figure}
    \centering
    \includegraphics[width=\linewidth]{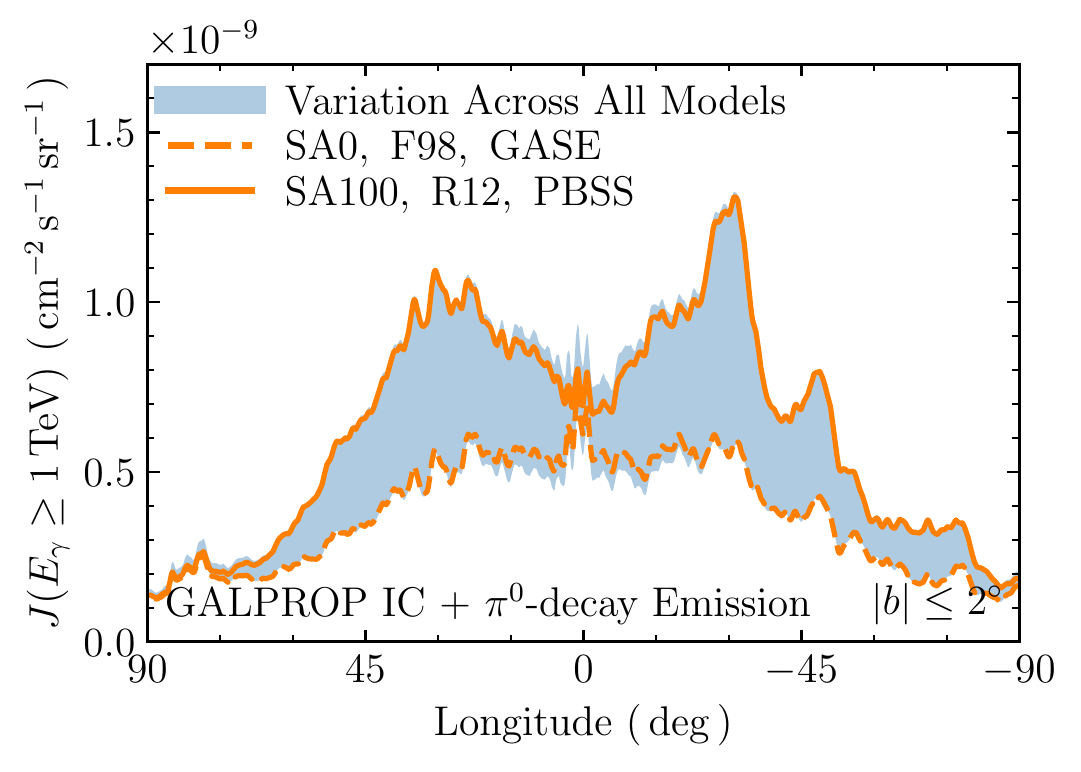}
    \caption{The envelope of the longitudinal profiles of the total average flux integrated above 1\,TeV for Galactic latitudes $|b|\leq 2^{\circ}$. Two individual \GP{} runs are shown, using the SA100, R12, and PBSS distributions~(solid), and SA0, F98, and GASE distributions~(dashed).}
    \label{fig:GALPROP source models and ISRFs}
\end{figure}

Fig.~\ref{fig:GALPROP source models and ISRFs} shows the envelope of longitudinal profiles at 1\,TeV of all the \GP{} runs across our model grid, i.e.~any combination of the source distribution, ISRF, and GMF.
Due to using a mix of distributions that include the spiral arms and those that do not~(e.g.~combining the SA0 with the PBSS distribution), no single \GP{} run describes the lower or upper bound of the envelope.
However, from the upper/lower bounds we infer that there is a $\pm 50$\% variation over the entire grid of \GP{} predictions for models that are consistent with the local CR data.
Changing the source distribution alone tends to alter the longitudinal profile most significantly toward the spiral arm tangents, where the variation can be up to $\sim$40\% when comparing SA0 to SA100.
Toward the inner Galaxy there are smaller differences, due to both the lower number of sources about the GC for distributions with spiral arms, and to the smoothing from the LOS integration over the long path lengths for these directions.
For similar reasons, changing the ISRF model impacts the results only modestly~(up to $\sim$4\%) around the central plane $|l| \leq 30^{\circ}$.
Meanwhile, changing the GMF model has the strongest effect around the central $|l| \leq 100^{\circ}$ where the impact can be up to $\sim$20\% variation.
We show overlaid in Fig.~\ref{fig:GALPROP source models and ISRFs} two individual results -- one with no spiral arms in the CR injection distribution, ISRF, or GMF~(i.e.~the SA0, F98, and GASE configuration), and the other having spiral arms in all three~(i.e.~the SA100, R12, and PBSS configuration).
The variation between the predictions is generally driven by the increasingly complex 3D structures for the respective components of the models that we have considered.%
\section{Comparing the HGPS to \GP}

The HGPS represents the most comprehensive survey of the Galactic plane at TeV energies to date.
Extraction of an estimate for the HGPS diffuse contribution requires accounting for the treatment of the CR background in the data. For this \hess{} used the adaptive ring method, with a set of exclusion zones that cover approximately two thirds of the Galactic plane, such that no \gray{} sources are considered as part of the CR background estimate. However, this may subtract the diffuse $\gamma$~rays that are captured in the background regions.
The HGPS observations also include both discrete and unresolved sources, neither of which are part of the diffuse emission.
We briefly describe how we address these issues below, with the full details given in Appendix~\ref{sect:HGPS}.

Our analysis of the HGPS uses the survey sky map with an integration/containment radius of $R_{c}=0.2^{\circ}$ such that we are as sensitive as possible to the TeV diffuse \gray{} emission.
We then use a sliding window method with parameters suitably optimised to reveal the longitudinal structure of the diffuse emission~(see Appendix~\ref{sect:HGPS}).
The sliding window is applied after the masking of known catalogued sources, which either subtracts the catalogued source component of the emission or excludes the catalogued source region from the analysis.
It is also necessary to account for the unresolved sources, i.e.~the ensemble of \gray{} sources below the detection threshold of \hess.
Although they are not individually resolved, these unresolved sources still contribute to the total observed Galactic emission, providing an extended low surface brightness contribution to the observed flux. As they are not part of the truly diffuse emission, their contribution must be subtracted.
We use estimates for the unresolved source component from~\citet{SteppaC.2020}~(SE20) and~\citet{CataldoM.2020}~(C20) who give relative contributions to the HGPS of 13\%--32\% and 60\%, respectively.
We combine them with the systematic uncertainty in the \hess{} fluxes to provide lower and upper limits for the diffuse \gray{} emission.

\begin{figure}
    \centering
    \includegraphics[width=\linewidth]{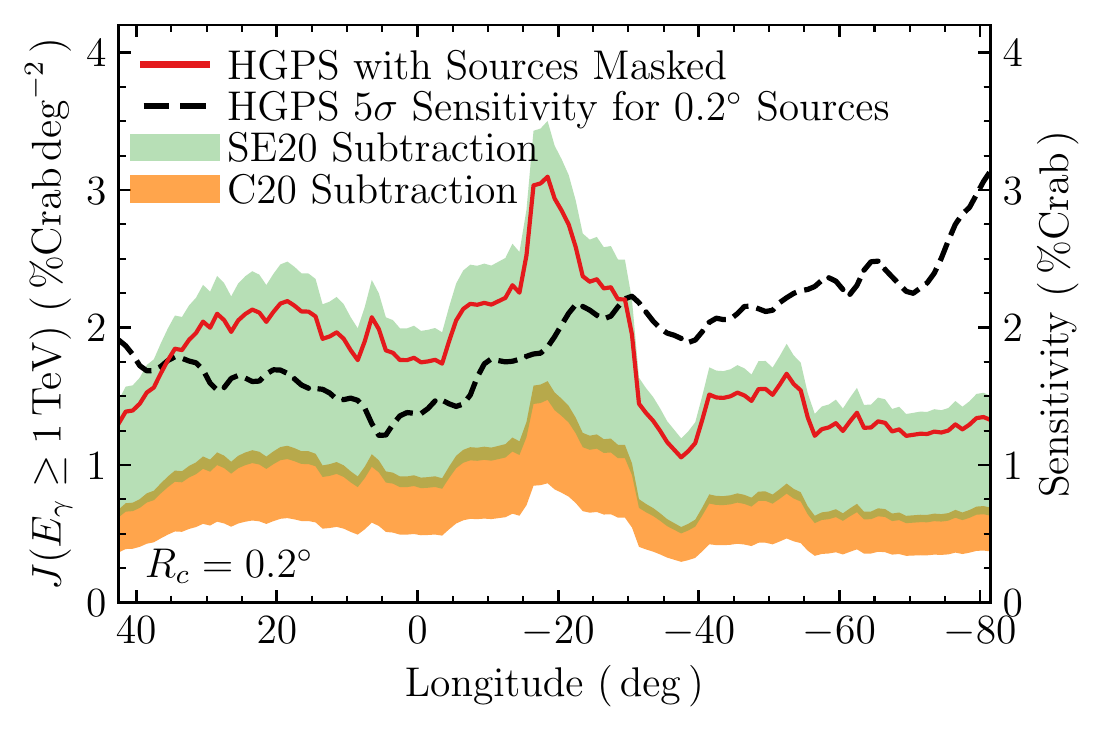}
    \caption{Longitudinal profile of the HGPS emission with catalogued sources subtracted~(red), shown after the sliding window has been applied for a containment radius of $R_{c}=0.2^{\circ}$ in units of \%Crab\,deg$^{-2}$. Also shown are the effects of subtracting the unresolved source fractions as estimated by~\citet{SteppaC.2020}~(SE20;~green) and~\citet{CataldoM.2020}~(C20;~orange). The $5\sigma$ sensitivity of the HGPS for $R_{c}=0.2^{\circ}$ is shown by the dashed black line in units of \%Crab.The analysis uses an averaging windows with a width of $\Delta w=15^{\circ}$ and height of $\Delta h=2.5^{\circ}$, centred at a latitude of $b_{0}=-0.25^{\circ}$ and spaced $\Delta s=1^{\circ}$ apart.}
    \label{fig:HGPS uncertainty}
\end{figure}

Fig.~\ref{fig:HGPS uncertainty} shows the source-subtracted flux obtained with our method, together with the bounding estimates employing the different unresolved source contributions.
This is shown together with the longitudinal profile of the HGPS sensitivity in units of \%Crab\footnote{Units of \%Crab are generally used in VHE \gray{} astronomy. We use the definition $J_{\mathrm{Crab}} (E \ge 1\,\mathrm{TeV})=2.26\times10^{-11}\,\mathrm{cm}^{-2}\,\mathrm{s}^{-1}$~\citep{2006A&A...457..899A}}.
There is a large-scale \gray{} emission along the plane, similar to that found previously by~\citet{AbramowskiA.2014}.
Its brightness varies over the survey area, with the lowest level $\sim$2--3 times below the $5\sigma$ point source sensitivity of the HGPS.

\begin{figure}
    \centering
    \includegraphics[width=0.47\textwidth]{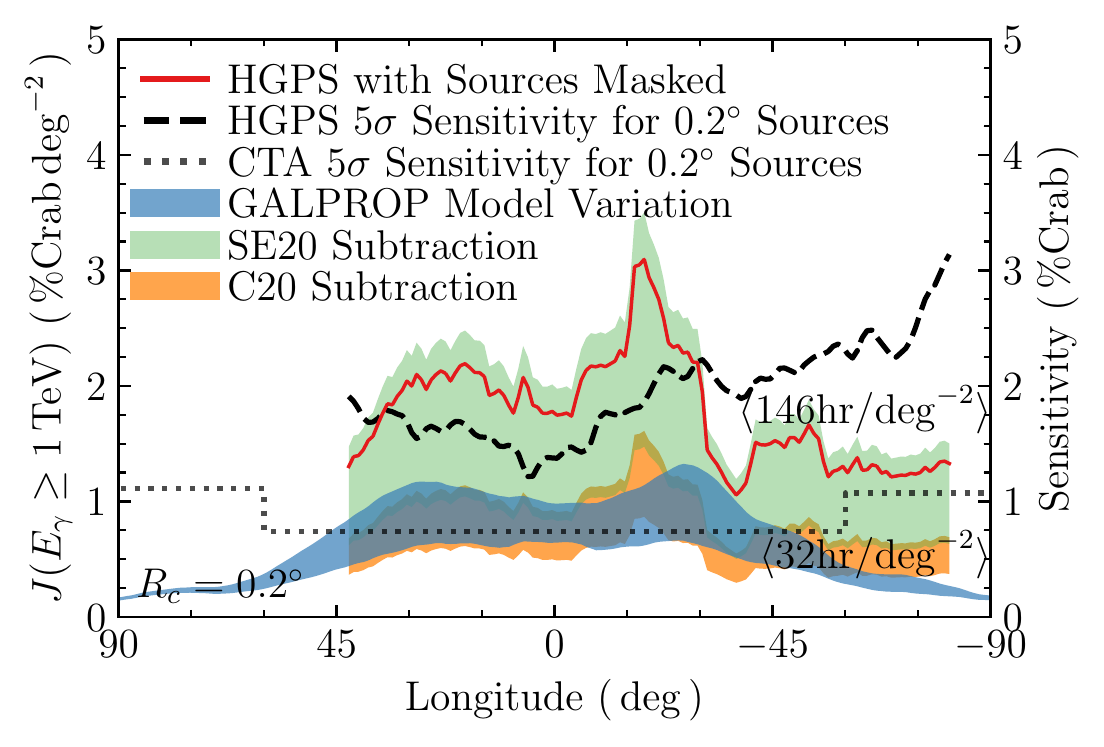}
    \caption{Longitudinal profiles integrated above 1\,TeV for the \GP{} variation~(blue) and the HGPS after catalogued sources are subtracted~(solid red), shown after the sliding window has been applied for a containment radius of $R_{c}=0.2^{\circ}$ in units of \%Crab\,deg$^{-2}$. Also shown are the results after subtracting the unresolved sources, with the estimates of the unresolved sources calculated from the estimates given by~\citet{SteppaC.2020}~(SE20;~green) and~\citet{CataldoM.2020}~(C20;~orange). The $5\sigma$ sensitivity is shown for the HGPS~(dashed black) and CTA GPS~(dotted black) when using a containment radius of $R_{c}=0.2^{\circ}$, both of which are shown in units of \%Crab. Averaging windows were applied to both the HGPS and the \GP{} profiles, and have a width of $\Delta w=15^{\circ}$ and height of $\Delta h=2.5^{\circ}$, centred at a latitude of $b_{0}=-0.25^{\circ}$ and spaced $\Delta s=1^{\circ}$ apart.}
    \label{fig:GALPROP vs HGPS with Sensitivities}
\end{figure}

We apply the same sliding window procedure to the \GP{} predictions over our model grid~(Section~\ref{sect:GALPROP}) to enable a comparison with the HGPS diffuse flux estimates.
Our \GP{} results were converted into units of \%Crab\,deg$^{-2}$ to allow comparisons to the HGPS sensitivity, which is given in units of \%Crab.
As \GP{} predicts only the diffuse interstellar emissions, the masking of catalogued and subtracting of the unresolved source component is not required.
Fig.~\ref{fig:GALPROP vs HGPS with Sensitivities} shows the comparison between the HGPS longitudinal profile for the range of the \GP{} predictions~(see Fig.~\ref{fig:GALPROP source models and ISRFs}) after the averaging window has been applied.
Also shown are the HGPS estimates with subtraction of the SE20 and C20 unresolved source fractions and HGPS sensitivity profile.

Our \GP-based models generally underpredict the large-scale TeV emission estimated from the HGPS.
This is due to the HGPS data containing a large number of unresolved sources that artificially inflate the large-scale emission to levels above the TeV diffuse emission.
Subtracting the unresolved sources from the large-scale emission via the estimates from~SE20 or~C20 provides a lower limit on the Galactic diffuse TeV emission.
Accounting for this contribution, the \GP{} predictions broadly agree with the diffuse \gray{} emission estimate obtained by subtracting an unresolved source fraction from~C20.
However, this is not true across the whole longitude range where, e.g.~around $l \sim -30^{\circ}$, we can find better agreement with estimated diffuse flux using the unresolved source fraction from~SE20.
It is therefore likely that the unresolved source contribution lies between those determined by~SE20 and~C20 across the region spanned by the survey, but it is not clear where either of the models is most applicable.

For Galactic longitudes $-80^\circ \leq l \leq -60^\circ$ the \GP{} predictions are below the lower-limit for the TeV diffuse emission.
However, this region of the Galactic plane is well below the $5\sigma$ sensitivity of the HGPS, making it difficult to draw any conclusions on the diffuse emission.
The disagreement could be due to poor statistics in the region, or that the \GP{} models are insufficiently optimised toward these directions.
\section{Discussion}

\subsection{Interstellar Emission Model Predictions into the TeV Energy Range}

Our work is the first systematic study using the \GP{} framework to predict the diffuse emissions into the TeV energy range for a grid of physical modelling configurations.
For the steady-state models that we have considered, generally the $\pi^0$-decay component produced by the CR nuclei interactions with the interstellar gas has only a small variation over the HGPS survey area~(Fig.~\ref{fig:IC and pion flux}).
Because the energy losses of these CRs are slow, they are distributed fairly smoothly across the MW with only mild imprints of the source density distribution.
However, this is not the case for the IC emissions.
Averaged over the HGPS survey area there is a much stronger variation due to the choice of source distribution.
The energy losses for the CR electrons are much more rapid than those for the nuclei.
The concentration of the ISRF and GMF energy densities about the spiral arms combines to affect the \gray{} intensities toward those regions, because they approximately coincide toward where the source densities with arms tend to also be enhanced.
Into the TeV energy range, the IC emissions have much higher sensitivity to the ISRF and GMF energy density distributions.

There are three composite source density distributions in our source density model grid~(both Galactic disc and spiral arm components: SA25, SA50, and SA75). For a fixed target density distribution~(i.e.~holding the ISRF/GMF constant), comparing the 1\,TeV \gray{} longitudinal profiles for these three composite source density distributions shows a variation of~$\sim$10\%.
When comparing all five source density distributions, we find a much larger variation up to~$\sim$50\%.
Altering the target densities as well as the source density distributions, as is done in Fig.~\ref{fig:GALPROP source models and ISRFs}, shows a variation up to~$\sim$100\%.
The effect of progressively more complex 3D structures for the source density and ISRF/GMF components leads to more variation for the TeV emissions from the CR electrons for the propagation model phenomenology that we have used in this paper.
However, the envelope of model predictions resulting from variations in the target densities is larger than that from solely changing the source density distribution.
While our current understanding of the spatial distributions of the ISM components that are the targets for producing the VHE emissions is incomplete, we have used models for them that are reasonably agreeing with data. The \GP{} predictions from our study are therefore representative of the likely uncertainty associated with current modelling for the VHE diffuse emissions.

We have used a grid of source density models where the normalisation is obtained by requiring consistency of the propagated CR intensity distributions with the data obtained at the solar system.
This is a well-motivated normalisation condition, given the considerable uncertainties associated with correctly defining CR source population characteristics.
However, there is an effective CR `horizon' beyond which additional components may be added without any noticeable effect when normalising to the local CR data.
For example, CRs injected about the GC likely have a negligible impact on the intensities at the solar system location.
Correspondingly, a source density localised about the inner Galaxy, e.g.~distributed similar to the stellar bulge/bar \citep{PorterT.2017,2019JCAP...09..042M}, will have its injected CR power as a free parameter.
Such additions localised about the inner Galaxy require tuning at other frequencies, e.g.~the GeV energy data, to correctly optimise parameters for use as a `background' in the VHE regime.
While the uncertainty band for the diffuse emissions that we estimated above is below the 5$\sigma$ significance level of the HGPS~(see Fig.~\ref{fig:GALPROP vs HGPS with Sensitivities}), such additional components have the possibility to also contribute toward the inner Galaxy. At the moment it is likely that a unique solution is not possible for precise estimation of the diffuse background for the HGPS data.

Our modelling has also shown that the IC emissions across the plane starts to become dominant over the $\pi^{0}$-decay above 10\,TeV~(Fig.~\ref{fig:diff. profile}).
The profiles for $\gtrsim$10\,TeV \gray{} energies display structure in the leptonic component that is due to the combination of the spatial distributions employed for our models.
However, at the higher energies~($\sim$100\,TeV) the situation becomes simpler, because the complications of the uncertainty in the ISRF distribution effectively disappear by the KN suppression of the optical and IR components.
Only the GMF and source density distributions are determining the structure visible in the longitudinal profiles for these energies.
Currently, the GMF models for the MW are poorly constrained, and are most commonly constructed by using rotation measures of extragalactic sources.
Given that the GMF structure is observable at 100\,TeV, the future VHE/UHE \gray{} observations provide an interesting possibility to constrain the GMF configurations.
However, \gray{} emission is created by the convolution of the source density distribution and the target density distributions. The alignment of the GMF and source density distribution is not complete, which is not surprising given that the different tracers for the Galactic spiral structure show offsets \citep[e.g.][]{2020ApJ...896...19V}.
Therefore using \gray{} observations to constrain the GMF will likely require \gray{} facilities many times more sensitive than what is currently available.

We note that for this paper we have only considered steady-state models.
Into the TeV energy range the time-dependent solutions, particularly for the CR electrons, are likely necessary \citep{PorterT.2019}.
For example, the \tibet{} data indicate that many of the $\gamma$~rays detected within their $25^\circ<l<100^\circ$, $|b|<5^\circ$ window \citep{AmenomoriM.2021} are not close on the sky to known VHE \gray{} sources.
The observational data from \tibet{} appears inconsistent with the simple models of \citet{LipariP.2018} and \citet{2019JCAP...12..050C}\footnote{These efforts do not use propagation codes, instead employing simple parametrizations for the CR spectra assuming steady-state conditions, and do not consider the 3D structure of the CR sources and ISM.}, and a discrete origin has been suggested \citep[e.g.][]{2021arXiv210402838D,2021ApJ...919...93F}.
Recent results from~\citet{VecchiottiV.2022} found that the discrepancy between models and the \tibet{} flux could be due to a population of unresolved PWNe, which would likely increase the leptonic component of the $\gtrsim 100$\,TeV $\gamma$~rays.
PWN halos have also been proposed as a large contributor to the unresolved sources for both \hess{} and HAWC~\citep{MartinP.2022}, further suggesting that electrons contribute to the diffuse emission.
The increasing contribution by the leptonic component with energy that we have shown in this paper, and its likely significant fluctuations, may help explain the lack of correlation between the PeV $\gamma$~rays observed by \tibet{} and neutrinos seen along the Galactic plane.
We also have results from HAWC~\citep{AbeysekaraA.2017} and LHAASO~\citep{Zhao:2021GJ} that show for energies $\gtrsim$10\,TeV the localisation of emissions about individual sources is stronger.
Incorporating time-evolution into the modelling seems to be necessary next step to accurately connect the GeV to 100s~TeV energies emissions that are presumably coming from a common origin.

\subsection{Application to CTA and Other TeV Facilities}

CTA will perform a Galactic plane survey~(GPS) over its first ten years of operation, which will cover Galactic latitudes in the range $b \leq |5^{\circ}|$ and all Galactic longitudes, averaging 32\,hours of observation per square degree~\citep{TheCTAOConsortium.2018}. The CTA GPS will have an energy range from 100\,GeV to 100\,TeV, with a resolution of around $3^{\prime}$ for energies above 1\,TeV. The $5\sigma$ point source sensitivity of CTA's proposed 10-year survey will reach 1.8\,mCrab for the inner $l \leq |60^{\circ}|$ and 3.8\,mCrab for the outer Galactic longitudes. For comparison, the HGPS has a point source sensitivity of 4--20\,mCrab between $40^{\circ} \leq l \leq 80^{\circ}$ with an average of 146 hours of observation per square degree. The sensitivity of the proposed CTA GPS after adjusting for a containment radius of $R_{c}=0.2^{\circ}$ is also shown in Fig.~\ref{fig:GALPROP vs HGPS with Sensitivities}.

As CTA is expected to detect and resolve on the order of a thousand of new sources~\citep{2011ExA....32..193A}, these exclusion regions will grow to encapsulate a significant fraction of the Galactic plane, if not the entire plane. Allowing nearby sources to contaminate the background, such as would be very likely along the Galactic plane without careful consideration, would significantly reduce the sensitivity of the observations~\citep{AmbrogiL.2016}.

CTA presents a leap forward in terms of the angular resolution it will achieve, with a PSF reaching approximately $1.8^{\prime}$ for \gray{} measurements at 1\,TeV.
The all-sky \hi{} and CO survey data that is used for building the neutral ISM model used in this work have resolutions of $27.5^{\prime}$ and $13.7^{\prime}$, respectively.
This is adequate for comparisons to the HGPS, but not to CTA's expected Galactic plane survey.
New high-resolution ISM surveys, such as Mopra~CO~\citep{2015PASA...32...20B, BraidingC.2018} in the Southern hemisphere, and Nobeyama~45m~CO~\citep{2015EAS....75..193M, 2017PASJ...69...78U} and THOR~\hi/OH~\citep{2016A&A...595A..32B, 2020A&A...634A..83W} in the Northern hemisphere have become available.
High-resolution ISM surveys that can observe the small, dense clumps at comparable or better angular resolution to CTA will become a necessary modelling ingredient for future high-resolution comparisons to its GPS. Due to the high density of these clumps it will also become necessary to adapt the diffusion coefficient to the density of the ISM gas~\citep[e.g.][]{GabiciS.2007,GabiciS.2009}.

From Fig.~\ref{fig:GALPROP vs HGPS with Sensitivities} we can see that CTA-South should be able to detect the TeV diffuse \gray{} emission to the 5$\sigma$ level given even our conservative estimates of the diffuse \gray{} emission predicted in this paper for the central $80^{\circ}$ of longitude~($|l| \leq 40^{\circ}$).
As has already been seen with the sensitive Fermi-LAT data at lower energies, it will therefore be necessary to develop accurate models for the diffuse emission to separate individual source characteristics from it.
This will be essential for resolving and detecting faint, extended TeV emissions such as those such as those expected from PWN haloes, and for probing complex morphological structures of TeV sources.
This will potentially allow the identification of currently unidentified sources in the HGPS and other current and future observations, such as those from CTA.

For the Galactic longitudes open to Northern-hemisphere observations, i.e.~$l>|50^{\circ}|$, our estimates state that the diffuse emission flux at 1\,TeV is below 5\,mCrab/deg$^{2}$. This emission is below the extended source sensitivity for CTA-North, which is planned to perform the deepest Galactic plane survey of any IACT in the Northern hemisphere at 1\,TeV~\citep{TheCTAOConsortium.2018}. It is therefore unlikely that any IACT in the Northern hemisphere, such as VERITAS~\citep{VERITAS_Sensitivity} and MAGIC~\citep{MAGIC_Sensitivity}, will observe the diffuse emission.
Ground-level particle detectors in the Northern hemisphere, such as LHAASO~\citep{LHAASO_Sensitivity}, HAWC~\citep{HAWC_Sensitivity}, and \tibet~\citep{TibetASg_Sensitivity}, are optimised for multi-TeV to PeV observations. This has allowed \tibet{} to successfully detect extended, presumably diffuse, emission at PeV energies~\citep{AmenomoriM.2021}.
However, as these particle detectors are optimised for multi-TeV to PeV observations, it is unlikely they will observe this diffuse emission around 1\,TeV within their first ten years of operation.
The Northern hemisphere detector ARGO-YBJ was constructed to be more sensitive at TeV energies, and has measured the diffuse emission at 1\,TeV after five years in the region~$25^{\circ}<l<100^{\circ}$.
The emission predicted by our \GP{} models is within the uncertainty band estimated for these data~\citep{BartoliB.2015}.

Observation of the Galactic diffuse emission below the PeV regime~(e.g.~around 1\,TeV) will likely remain difficult with the current and the next generation of \gray{} observatories. For the Southern hemisphere, the planned ground-level particle detector SWGO~\citep{SWGO_whitepaper} has the possibility of observing the Galactic diffuse emission at 1\,TeV around the GC within the first ten years of operation, complementing the higher-resolution CTA-South observations.%
\section{Summary}

We have made a systematic comparison of the predicted diffuse TeV \gray{} emission over a grid of steady-state models generated using the \GP{} framework.
For the models that we consider, the variations in source density, ISRF, and GMF distributions can have a significant effect on the predicted IC emissions at VHEs.
The magnitude of the variation on the IC emissions around 1\,TeV toward the GC can be $\sim$50--60\%.
This increases to $\sim$100\% for $\sim$100\,TeV energies.
The influence of differing ISRF distributions is most important for energies $\lesssim$1\,TeV, modest for $\sim$1--10\,TeV, and vanishes $\gtrsim$40--50\,TeV due to the KN suppression that removes all but the CMB as a target photon field for Compton scattering processes at the highest energies.
For energies $\gtrsim$~20--30\,TeV, the CR source density and GMF distributions have the most significant impact on the predicted \gray{} emissions.
Generally the energy losses at VHEs for the CR electrons are most strongly influenced by the GMF distribution.
The structure of the GMF is imprinted on the CR electron intensity distribution and hence the VHE diffuse emissions produced by them.
This is seen in the IC emissions, with the GMF structure becoming more pronounced as energy increases into the $\sim$100\,TeV range.

Towards the inner Galaxy~($|l|<40^{\circ}, \ |b|<2^{\circ}$) we found that the leptonic component of the diffuse \gray{} emission begins to become an important consideration around 1\,TeV, with the hadronic component dominating for most of the Galactic plane. At 10\,TeV the leptonic emission begins to dominate over the hadronic emission for the inner Galaxy, while the leptonic component is approximately equal in magnitude to the hadronic emission for the region~($|l|>40^{\circ}, \ |b|<2^{\circ}$).
For the $\sim$100\,TeV energy range, our results show that the IC emission dominates the total diffuse \gray{} emission across the entire plane for all models that we considered.
With the GMF structure imprinted on the IC emission in this energy regime, observations of the diffuse \gray{} emission are in a unique position to constrain the GMF structure.
Furthermore, our \GP{} predictions and the importance of leptonic emissions for \gray{} energies $\gtrsim$100\,TeV may provide some insight on comparisons between the PeV diffuse \gray{} emission and neutrino fluxes.
Due to how rapidly electrons in the 100\,TeV to 1\,PeV energy range lose their energy, our results motivate the need for a time-dependent solution for future work at these energies.

Our \GP{} predictions of the diffuse \gray{} emission match lower limits on the diffuse TeV emission inferred from the Galactic plane survey carried out by the \hess{} collaboration, after subtracting the unresolved source fraction estimates calculated from~SE20~\citep{SteppaC.2020} and~C20~\citep{CataldoM.2020}.
Because the \GP{} predictions overlap the lower limits of the TeV diffuse emission inferred from the HGPS after subtracting catalogued sources and estimates of the unresolved source contribution, further optimisation of the models may not yield a meaningful impact on analysis for these data.
However, the brightness of the \GP{} predictions indicates that it will be necessary to better optimise the models for the more sensitive proposed CTA GPS.
The need for better, physically-motivated diffuse emission models at VHEs is further reinforced now that both LHAASO and \tibet~\citep{Zhao:2021GJ, AmenomoriM.2021} have observed sources of PeV $\gamma$~rays within the MW.%

\section*{Acknowledgements}

This research is supported by an Australian Government Research Training Program Scholarship. \GP{} development is partially funded via NASA grants NNX17AB48G, 80NSSC22K0718, and 80NSSC22K0477. Some of the results in this paper have been derived using the \verb|HEALPix|~\citep{HEALPix} and \verb|Astropy|~\citep{astropy:2013, astropy:2018} packages. This work was supported with supercomputing resources provided by the Phoenix HPC service at the University of Adelaide, and we want to thank F.~Voisin in particular for his many hours spent configuring the HPC service to work efficiently with \GP.

\section*{Data Availability}

The \GP{} configuration files can be found via the \GP{} website: \url{https://galprop.stanford.edu/}, and the HGPS results can be found via the \hess{} Galactic plane survey~\citep{HESS_GPS.2018} website: \url{https://www.mpi-hd.mpg.de/hfm/HESS/hgps/}.



\bibliographystyle{mnras}
\bibliography{ref.bib} 




\appendix

\section{Transforming the HGPS for Comparisons to \GP} \label{sect:HGPS}

The HGPS comprises about 2700~hours of observations spanning over 10~years.
It covers almost 180~degrees of longitude, ranging from $l=270^{\circ}$ to $l=65^{\circ}$, and spans over the latitudes $b \leq |5^{\circ}|$ with a resolution of $4.8^{\prime}$.
The survey has a variable point-source sensitivity above 1\,TeV being as low as just under $\sim$0.5\% of the Crab flux~(4\,mCrab) for the GC, worsening to 20\,mCrab as exposure time decreases at the edges of the survey.
For this paper we used the flux maps available at \url{https://www.mpi-hd.mpg.de/hfm/HESS/hgps/} from~\citet{HESS_GPS.2018}.

The HGPS contains emission from discrete sources and unresolved sources on top of the diffuse \gray{} emission.
The data also has the CR background influencing the measurements, and a variable exposure.
To compare the HGPS to our \GP{} predictions, the HGPS needs to be transformed into a form more similar to the skymaps that \GP{} creates.
In this appendix we detail how the HGPS transformation is made and detail our analysis method.

\begin{figure}
    \centering
    \includegraphics[width=0.48\textwidth]{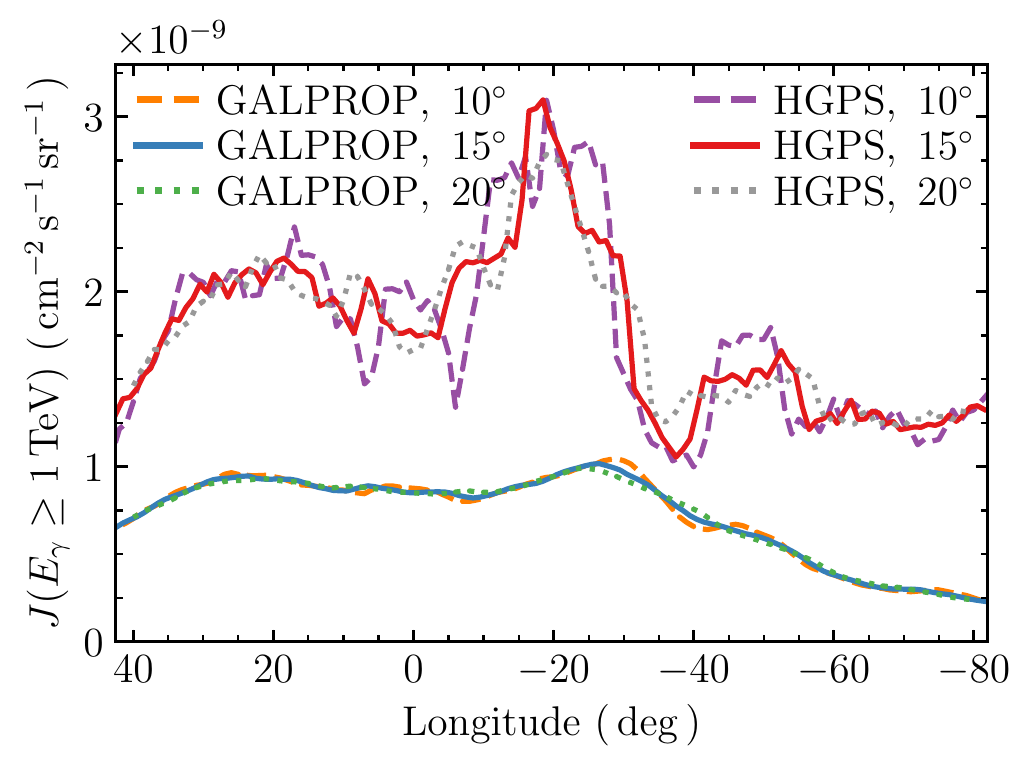}
    \caption{Longitudinal profiles of the $R_{c}=0.2^{\circ}$ HGPS and one of the \GP{} simulations integrated above 1\,TeV. The width of the sliding window, $\Delta w$, is varied from $10^{\circ}$ to $20^{\circ}$.}
    \label{fig:window width trials}
\end{figure}

A requirement of our analysis is that it should not be sensitive to the method used, and needs to be applied in a fair and representative way to both the HGPS observations and our \GP{} predictions. For this we follow the analysis used by \citet{HESS_GPS.2018} and perform a sliding window analysis.
We selected the width of the window,~$\Delta w=15^{\circ}$, such that the integration area was as large as possible while not over-smoothing large-scale features such as the Galactic arms, and such that the results were robust to changes of~$\Delta w$ on the order of 30\%. The sensitivity of the HGPS and \GP{} results to changing the window width is shown in Fig.~\ref{fig:window width trials}. 
Across most Galactic longitudes, the HGPS is most sensitive for $-1.5^{\circ} \leq b \leq 1.0^{\circ}$ as most of the observation time was spent in this region. These latitudes have a higher statistical significance, with latitudes outside this range having a significance below $5\sigma$ across wide areas of the Galactic plane. To ensure statistically significant data is used in our analyses, we chose the latitude bounds of the window to be defined by the height~$\Delta h=2.5^{\circ}$ and the central latitude $b_{0}=-0.25^{\circ}$. Finally, we chose the spacing between the windows, $\Delta s=1.0^{\circ}$, such that variation in the diffuse emission between \gray{} sources could be observed, as approximately~80\% of the \gray{} sources in the HGPS are separated by more than a degree in longitude.

\begin{figure}
    \centering
    \includegraphics[width=0.48\textwidth]{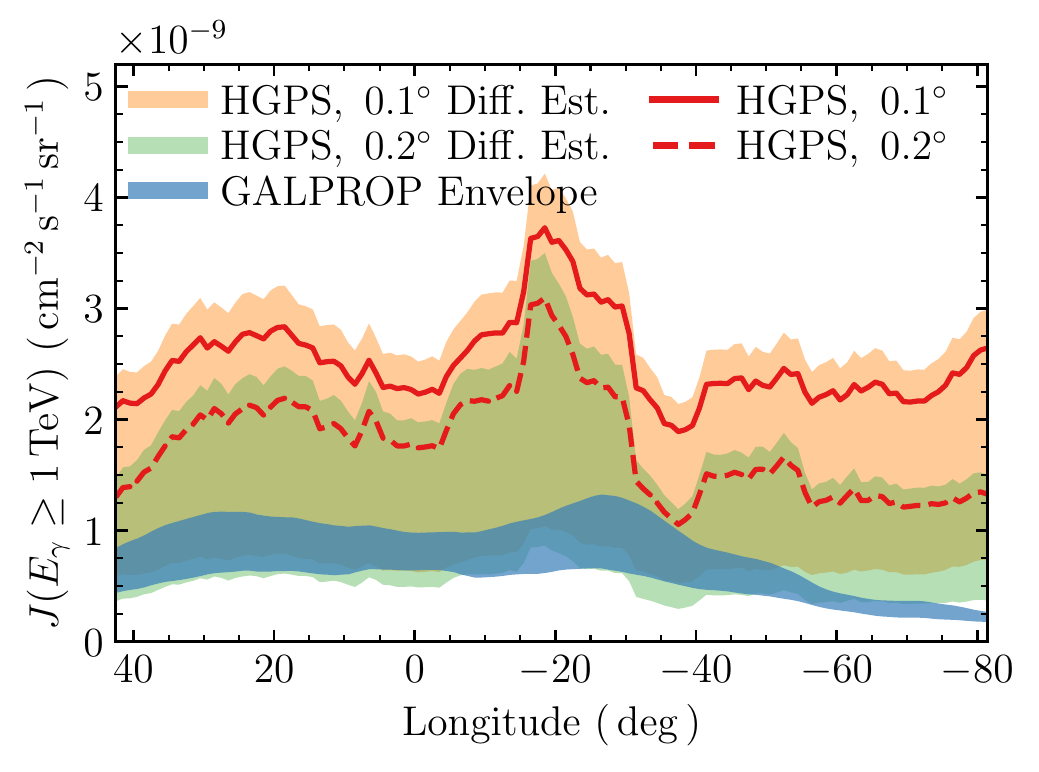}
    \caption{Longitudinal profiles of the HGPS and the envelope of the \GP{} results integrated above 1\,TeV. The beam size used in creating the map, $R_{c}$, is equal to $0.1^{\circ}$ or $0.2^{\circ}$. The unresolved source estimates from~\citet{SteppaC.2020} and~\citet{CataldoM.2020} have been combined into a single shaded band for each integration radii.}
    \label{fig:beam size trials}
\end{figure}

The HGPS is well suited for resolving smaller sources that are approximately $0.1^{\circ}$ or $0.2^{\circ}$ in radius as \hess{} uses an integration radius~($R_{c}$) of either~$0.1^{\circ}$ or~$0.2^{\circ}$.
However, as we expect the diffuse emission to vary on scales greater than~$0.2^{\circ}$, it is likely supressed by both integration radii.
Because this effect can conceal the large-scale emission, any estimate of the diffuse emission obtained from the HGPS data using $R_{c}=0.2^{\circ}$ must be regarded as a lower limit.
Fig.~\ref{fig:beam size trials} shows the sensitivity of the analysis to changes in $R_{c}$. \GP{} doesn't have a beam, and there was no impact when applying an artificial beam to \GP{} as the results are smoothly varying.

\begin{figure}
    \centering
    \includegraphics[width=\linewidth]{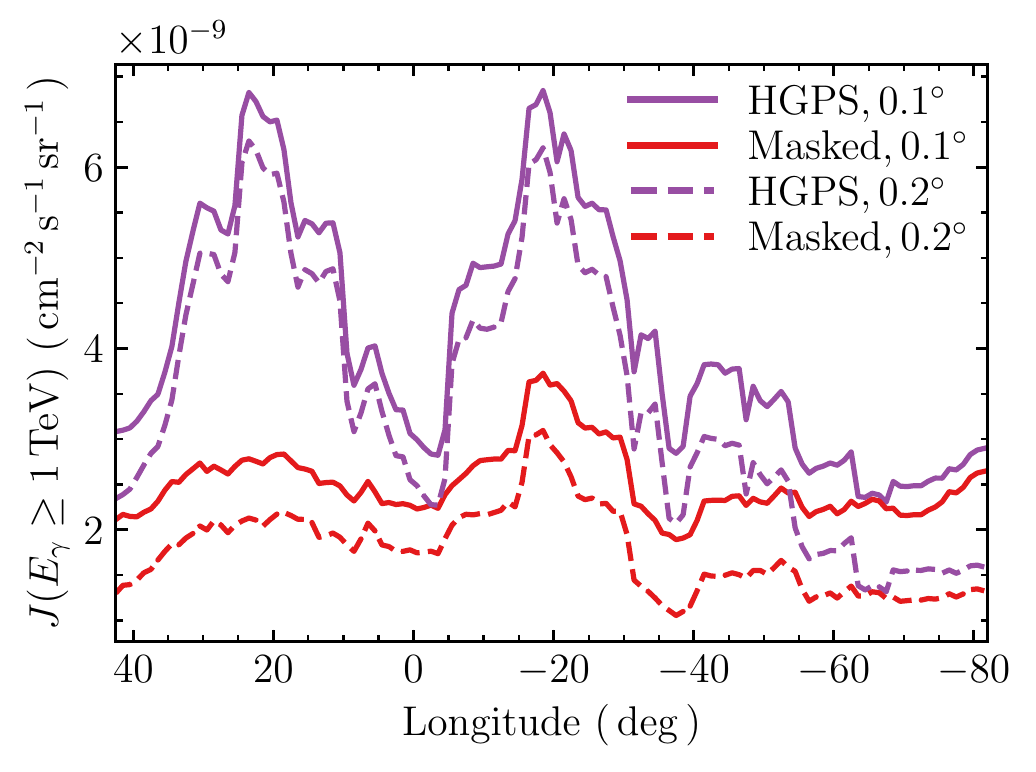}
    \caption{Longitudinal profiles of the average flux in the HGPS measurement. Shown are the HGPS results~(purple), and the HGPS results with the catalogued sources masked from the image~(red), with the two integration radii being $R_{c}=0.1^{\circ}$~(solid) and $R_{c}=0.2^{\circ}$~(dashed). Averaging windows have a width $\Delta w=15^{\circ}$ and height $\Delta h=2.5^{\circ}$, centred at a latitude of $b_{0}=-0.25^{\circ}$ and spaced $\Delta s=1.0^{\circ}$ apart.}
    \label{fig:HGPS source mask}
\end{figure}

\begin{figure*}
    \centering
    \includegraphics[width=\linewidth]{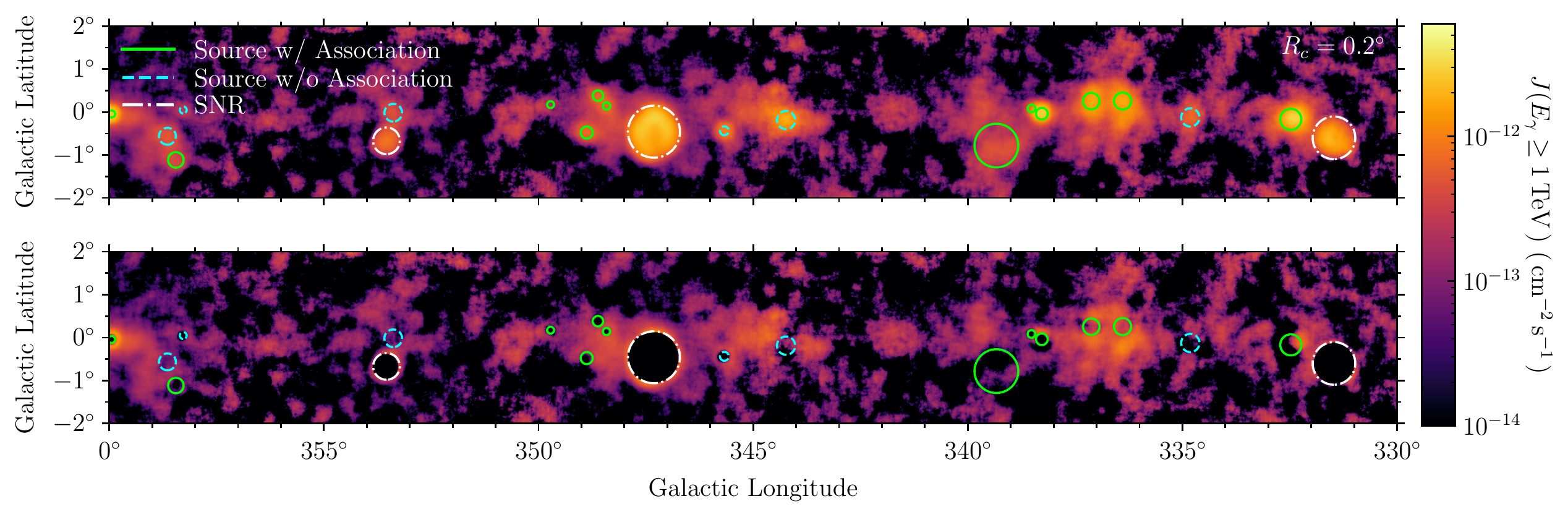}
    \caption{HGPS flux in cm$^{-2}$\,s$^{-1}$ between $l=330^{\circ}$ and $l=0^{\circ}$ for the $R_{c}=0.2^{\circ}$ map, comparing before~(top panel) and after~(bottom panel) sources are masked from the image. Sources with a known association are shown with a solid green circle, sources without an association are shown with a dashed cyan circle, and SNRs are shown with a white dash-dot circle.}
    \label{fig:mask demo}
\end{figure*}

The HGPS has catalogued a total of 78~sources in TeV $\gamma$~rays. These sources represent localised regions in \gray{} flux from local particle accelerators, and are not modelled by \GP. As the local source emission is not a part of the Galactic diffuse emission, we must mask these sources from the HGPS. The methods outlined by~\citet{HESS_GPS.2018} are followed here, with complex morphological structures such as SNRs being cut out from the image and excluded from our analyses. Simple morphological structures are assumed to have Gaussian profiles, with their surface brightness~($S_{\mathrm{Gauss}}$, equation~\eqref{eq:S_Gauss}) being subtracted from the image.
This subtracts only the source component of the flux from the image:

\begin{eqnarray}
    S_{\mathrm{Gauss}}(r|\phi, \sigma) &= \phi \frac{1}{2 \pi \sigma^{2}} \exp{\left( - \frac{r^{2}}{2\sigma^{2}} \right)} \label{eq:S_Gauss}
\end{eqnarray}

\noindent
where $\phi$ is the spatially integrated flux, and $\sigma$ is the radius of the source. The offset $r=\sqrt{(l-l_{0})^{2} + (b-b_{0})^{2}}$ is the radial distance from the source located at the Galactic coordinates~($l_{0},~b_{0}$).
A comparison between the HGPS with no source masking and the HGPS with all sources masked is shown in Fig.~\ref{fig:HGPS source mask}, where the results for all sources masked represents the residual emission along the Galactic plane on the scale of $R_{c}$, after the contribution from catalogued sources has been subtracted.
Also seen in Fig.~\ref{fig:HGPS source mask} is that the catalogued sources within the HGPS contribute a significant fraction of the measured \gray{} flux along the Galactic plane. A demonstration of the source masking can be seen for an arbitrary region in the Galactic plane in Fig.~\ref{fig:mask demo}.

\hess{} is unable to detect sources below a certain \gray{} luminosity, having a point-source horizon between 5--14\,kpc for $10^{34}\,\mathrm{erg}\,\mathrm{s}^{-1}$, decreasing to a horizon of 1--4\,kpc for point-sources with a flux of $10^{33}\,\mathrm{erg}\,\mathrm{s}^{-1}$. The HGPS horizon also decreases with an increasing source size. The sensitivity, given by the minimum flux~($F_{\mathrm{min}}$) that can be observed, is shown in equation~\eqref{eq:source size sensitivity}:
\begin{eqnarray}
    F_{\mathrm{min}}(\sigma_{\mathrm{source}}) &\propto \sqrt{\sigma_{\mathrm{source}}^{2} + \sigma_{\mathrm{PSF}}^{2}} \label{eq:source size sensitivity}
\end{eqnarray}

\noindent
where $\sigma_{\mathrm{PSF}}=0.08^{\circ}$ is the point spread function of \hess{} and $\sigma_{\mathrm{source}}$ is the source size.
Due to the sensitivity of \hess{} worsening for extended sources, and as the HGPS source horizon does not cover the entire Galaxy, there are hundreds of TeV \gray{} sources within the MW that have not yet been resolved by \hess~\citep{2011ExA....32..193A, HESS_GPS.2018}. An estimate of the unresolved source contribution from~\citet{SteppaC.2020} suggests that they contribute 13\%--32\% to the diffuse measurement by comparing the total flux observed in the HGPS to the total TeV luminosity produced by a distribution of sources with an estimate of the total luminosity of TeV sources based on PWNe distributions. Another estimate from~\citet{CataldoM.2020} states that the fraction of unresolved sources could be as large as 60\% by using known SNR and PWNe spatial distributions and a luminosity distribution to calculate the number of detectable sources as well as the total \gray{} flux that all the sources would produce. The result of this subtraction is shown in Fig.~\ref{fig:HGPS uncertainty}.


\bsp	
\label{lastpage}
\end{document}